\documentclass{aa}
\usepackage{graphicx}
\usepackage{natbib}
\bibpunct{(}{)}{;}{a}{}{,}
\begin{document}
   \title{Photometry and membership for low mass stars in the
young open cluster NGC 2516}

   \author{R. D. Jeffries
          \inst{1}\fnmsep\thanks{Visiting astronomer at the Cerro
          Tololo Interamerican Observatory, operated by the Association
          of Universities for Research in Astronomy, Inc., under
          contract to the National Science Foundation. email: rdj@astro.keele.ac.uk}
          \and
          M. R. Thurston\inst{2}
	  \and
          N. C. Hambly\inst{3}
          }

   \offprints{R. D. Jeffries}

   \institute{Department of Physics, Keele University, Keele, Staffordshire,
ST5 5BG, UK
         \and
School of Physics and Astronomy, 
University of Birmingham, Edgbaston, Birmingham, B15 2TT,
UK
\and
Institute for Astronomy, University of Edinburgh, Blackford
Hill, Edinburgh, EH9 3HJ, UK             }

   \date{Received; accepted}

   \abstract{
We present the results of a 0.86 square degree CCD photometric survey
of the open cluster NGC 2516, which has an age of about 150\,Myr and
may have a much lower metallicity than the similarly-aged Pleiades. Our
$BVI_{c}$ survey of cluster members is complete to $V\simeq20$ and is
used to select a preliminary catalogue of 1254 low mass
($0.2<M<2.0M_{\sun}$) cluster candidates, of which about 70--80 percent
are expected to be genuine.  After applying corrections for
contamination by non-members and adding data for higher mass stars from
the literature, we investigate the cluster binarity, luminosity and
mass function, mass segregation and total mass. We find a binary
fraction of $26\pm5$ percent, for A to M-type systems with mass ratios
between 0.6 and 1, which is very similar to the Pleiades.  The mass
function is metallicity and evolutionary-model dependent, but
consistent with a Salpeter-like law ($dN/d\log M\propto M^{-\alpha}$,
$\alpha=+1.47\pm0.11$ or $\alpha=+1.67\pm0.11$ for the solar and
half-solar metallicity models of Siess et al. (2000), and
$\alpha=+1.58\pm0.10$ for the solar metallicity models of
D'Antona \& Mazzitelli (1997)), for $0.7<M<3.0M_{\sun}$.  At lower masses
($0.3<M<0.7M_{\sun}$) there is a sharp fall in the mass function, with
$\alpha=-0.75\pm0.20$ or $\alpha=-0.49\pm0.13$ (for the solar and
half-solar metallicity models of Siess et al.), and
$\alpha=-1.00\pm0.18$ (for the solar metallicity models of D'Antona \&
Mazzitelli). The true stellar mass function might have $\alpha$ values up to 0.4
larger if account were taken of low mass stars in
unresolved binary systems with mass ratios less than 0.6. The falling mass
function of NGC 2516 at lower masses seems inconsistent with the much
flatter mass functions derived from comparable data in the Pleiades and
field populations.  This deficit of lower mass, fainter stars is also
seen in the observed luminosity function.  We rule out incompleteness
as the cause of this discrepancy, but demonstrate that mass segregation
is clearly present in NGC 2516, with more than half the low-mass
($<0.6M_{\sun}$) stars likely to lie outside our survey area, but the
vast majority of high-mass ($>1.5M_{\sun}$) stars included.  Taking
this into account, it is probable that the whole-cluster mass functions
for NGC 2516 and the Pleiades are similar down to 0.3$M_{\sun}$. The
mass of NGC 2516 stars with $M>0.3M_{\sun}$ inside our survey is
$950-1200M_{\sun}$, depending on metallicity and what corrections are
applied for unresolved binarity. Correcting for mass segregation
increases this to $\sim1240-1560M_{\sun}$, about twice the total mass
of the Pleiades. If NGC 2516 and the Pleiades do have similar mass
functions, then less massive stars and brown dwarfs contribute about a
further 15 percent to the mass of NGC 2516 and we predict a cluster
population of about 360--440 brown dwarfs with $0.030<M<0.075M_{\sun}$.
   \keywords{open clusters and associations: individual: NGC 2516 -- stars: mass
function, luminosity function -- binaries: general
               }
   }

\titlerunning{Photometry in NGC 2516}

   \maketitle

\section{Introduction}

Observations of co-eval stars in open clusters play a vital r\^{o}le in
investigating the low-mass stellar initial mass function and
defining the physical processes that drive the evolution of rotation,
magnetic activity and photospheric light element abundances in cool
stars with convective envelopes. The ability to study samples
with common, well determined, ages, distances and compositions is the
key to their usefulness.  A careful census and identification of
cluster members is usually a prerequisite of all such studies: (a) in
order to prevent contamination of sample properties by interloping
non-members that have different properties to the cluster stars and (b)
to select samples of cluster stars that are unbiased with respect to
the properties which are under investigation (\emph{e.g.} magnetic
activity).

Prior to 1990, most work in this area focussed on the nearby, well
studied clusters such as the \object{Hyades} and \object{Pleiades}, for which there was
ample pre-existing photometry and proper motion information. However,
in the last decade it has been realised that a wider range of clusters
need to be studied in detail. The reasons for this are:
\begin{enumerate}
\item To properly sample a wide age range, from clusters just emerging from
their parent molecular clouds at 10\,Myr, through mature clusters like
the Hyades at 600\,Myr to the rarer old clusters such as M67 with ages
of a few Gyr.
\item Assuming that the properties of stars in one cluster are 
representative of all such stars at a similar age is precarious and
neglects the possible influence of composition differences or birth
conditions on behaviour in later life.
\item To investigate dynamical effects such as mass segregation 
on cluster evolution and the possible universality of properties such 
as binary fractions and initial mass functions requires accurate
membership studies in clusters with a wide range of ages and stellar
densities.
\end{enumerate}

In the last few years the rich, southern Galactic open cluster, \object{NGC 2516}
(RA$=07^{h}58^{m}$, Dec$=-60^{d}45^{m}$, $l=274^{\circ}$, $b=-16^{\circ}$),
has become a key object in the study of low mass stars.
\citet{meynet93} give an age of 141\,Myr, in
comparison with ages for the better studied Pleiades and \object{$\alpha$ Per}
clusters of 100\,Myr and 52\,Myr respectively.  Photometric studies
of early-type cluster members 
were presented by \citet{cox55}, \citet{eggen72} and \citet{dachs89}, 
yielding a mean reddening estimate E(B-V)=0.12 and a distance of about
400\,pc.  A major X-ray study
was undertaken using the \emph{ROSAT} satellite by
\citet*{jeffriesn251697}.  
159 X-ray sources were identified within a 20 arcmin
radius of the cluster centre, 
65 of which could be identified as photometric members of the
cluster with V$<15$, the majority being cool, coronally active
stars. Using these stars, and comparing with other clusters, \citet{jeffriesn251697}
showed that NGC 2516 has a distance of about 390\,pc and
a $U-B$ excess that is best explained if its metallicity is substantially
less than the Pleiades.  A model dependent metallicity of
[Fe/H]$=-0.32\pm0.06$ was derived, in agreement with an earlier study
by \citet{cameron85b}. A sub-solar metallicity was also found using an
independent $B-V$ vs $V-I$ photometric technique by
\citet*{jeffriesn251698} and \citet{pinsonneault00}, but has yet to be
confirmed spectroscopically.

It is this possible low metallicity, together with its richness that makes NGC
2516 so interesting. There are plausible reasons why dynamo generated
activity, rotational spindown and light element depletion could be
profoundly affected by a low photospheric metallicity and thus
differing convection zone properties.  The compact size and numerous
cluster members make NGC 2516 an ideal target for fibre spectroscopy
and for X-ray satellites such as \emph{XMM} and \emph{Chandra}, which have limited
fields of view. Recently, new X-ray studies by the \emph{ROSAT} high
resolution imager \citep{micela00},
\emph{Chandra} \citep{harnden01} and \emph{XMM-Newton} \citep{sciortino01} have devoted
considerable amounts of time to observing NGC 2516. 
These more recent studies have used the photometric
catalogue described in this paper to identify cluster X-ray sources.

So far, systematic membership studies in NGC 2516 have been limited to
fairly bright stars or small areas. An unbiased, but severely
incomplete list of cluster members down to about $V=15$ is presented by
\citet{jeffriesn251697}, based upon position in the $V$ vs $B-V$
colour-magnitude diagram (CMD). \citet*{hawley99} present photometry
and low resolution spectroscopy in the cluster down to much fainter 
limits, though over a very small area. \citet{jeffriesn251698} present high
resolution spectroscopy of photometrically selected candidates between
$11<V<14.5$, establishing definite membership through radial velocity
measurements as well as lithium abundances. This work on brighter
objects has been continued recently by \citet{terndrup01}.

In this paper we present a catalogue of CCD photometry and astrometry
down to faint magnitudes ($V=20$) and over a wide area (0.86
square degrees) in NGC 2516. This catalogue will be an invaluable tool
for selecting unbiased samples of F, G, K and M stars for further
study, interpreting X-ray observations and studying the dynamical state
and mass function of NGC 2516. A preliminary version of this work
appeared in \citet{thurston99}. We will restrict ourselves in this paper
to a description of the data, the construction of a photometric
catalogue and a membership classification on the basis of
this photometry. Section 2 outlines the collection and reduction of
the photometric data, and deals with the astrometric calibration,
catalogue completeness and comparison with the previous
literature. Section 3
describes a method for selecting candidate cluster members, which
will need to be refined as more observational data become
available. Section 4 uses this catalogue to look at the
luminosity and mass function of NGC 2516 and search for evidence of mass segregation.
The catalogue itself can be obtained in electronic format from the
Centre de Donn\'ees astronomiques de Strasbourg.

\section{Observations}

\begin{figure}
\vspace*{9cm}
\includegraphics{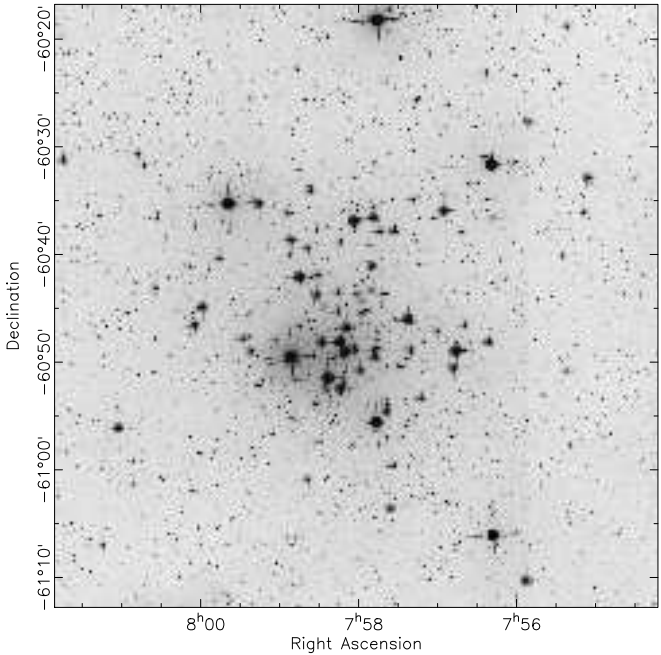}
\caption{An optical picture from the Space Telescope Science Institute
digitized sky survey showing the region of sky included in our CCD photometry.}
\label{n2516opt}
\end{figure}

The data used in this paper were collected at the 0.9-m telescope of
the Cerro Tololo Interamerican Observatory (CTIO) on the nights of 8 9 and 12
January 1995 (lunar phase 0.52-0.88). A CCD photometric survey was
completed over an approximate 0.86 square degrees, centred upon RA
$=07^{h}58^{m}02.8^{s}$, Dec$=-60^{d}44^{m}54^{s}$ (J2000.0), which
comprised of 25 overlapping frames of 13.5x13.5 square arcminutes,
taken with a 2048x2048 pixel Tektronix CCD at the Cassegrain focus. The
centres of the fields were separated by 10.5 arcminutes, leading to a
square survey area, $55^{m}36^{s}$ on a side (see Fig.~\ref{n2516opt}).
A series of short (30s, 15s, 15s) and long (200s, 100s, 100s) exposures
were taken through a set of Harris $BVI$ filters respectively. In
addition to these target exposures, on each night a set of high quality
twilight flat fields were obtained through each filter and many Landolt
(1992) standard fields were observed over the entire range of airmass
for which the cluster data were taken.

All of the frames were bias subtracted and flat-fielded with standard
methods using the Starlink {\sc figaro} package \citep{shortridge99}. The only
complications were that the CCD was read out using two amplifiers with
slightly different bias, gain and readout noise characteristics. This
was dealt with by reducing the two halves separately and then
correcting for the gain ratio by requiring that the flat fields were
continuous across the amplifier boundary. The flat-fielding was tested
by flat-fielding median stacked night sky exposures and found to
successfully remove most of the structure in the sky at a level better
than a few tenths of one percent, apart from several CCD cosmetic defects
(including a number of bad columns) and a narrow (10 arcsecond) strip
around the outside of the field.  A mask was constructed so that these
defective pixels were not considered in the subsequent reduction.

\subsection{Photometry}

\begin{figure*}
\vspace*{22cm}
\includegraphics{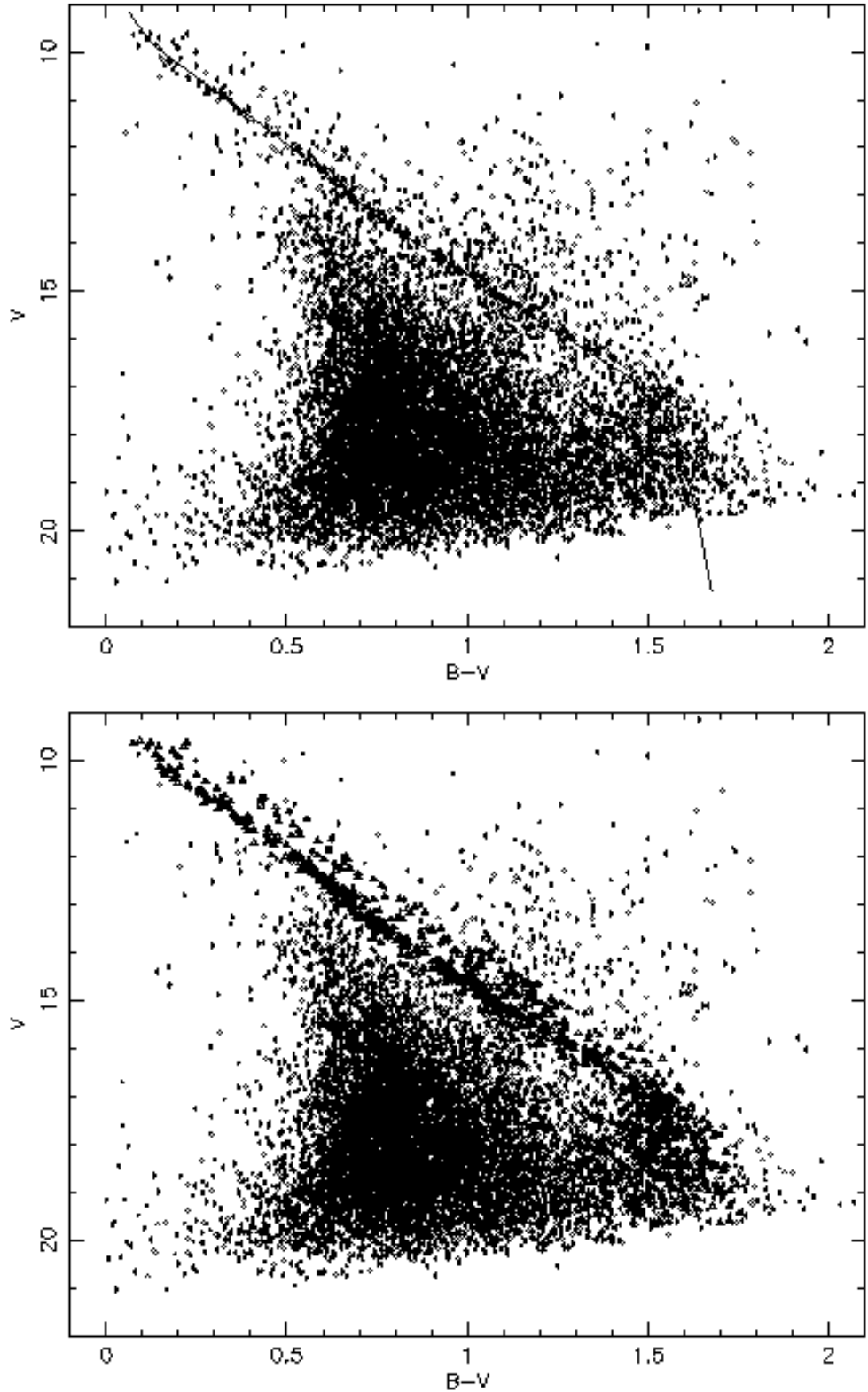}
\caption{Top: The $V$ vs $B-V$ CMD for our survey. The solid line shows
the \citet{siess00} solar metallicity 
isochrone we use to select members of the cluster (see
Sect.~\ref{isochrone}). Bottom: Objects selected as cluster members
(see Sect.~\ref{membership})
are shown as triangles (where both $B-V$ and $V-I_\mathrm{c}$ were
available) or squares (where only $B-V$ was available -- for just
5 objects).}
\label{bv}
\end{figure*}
\begin{figure*}
\vspace*{22cm}
\includegraphics{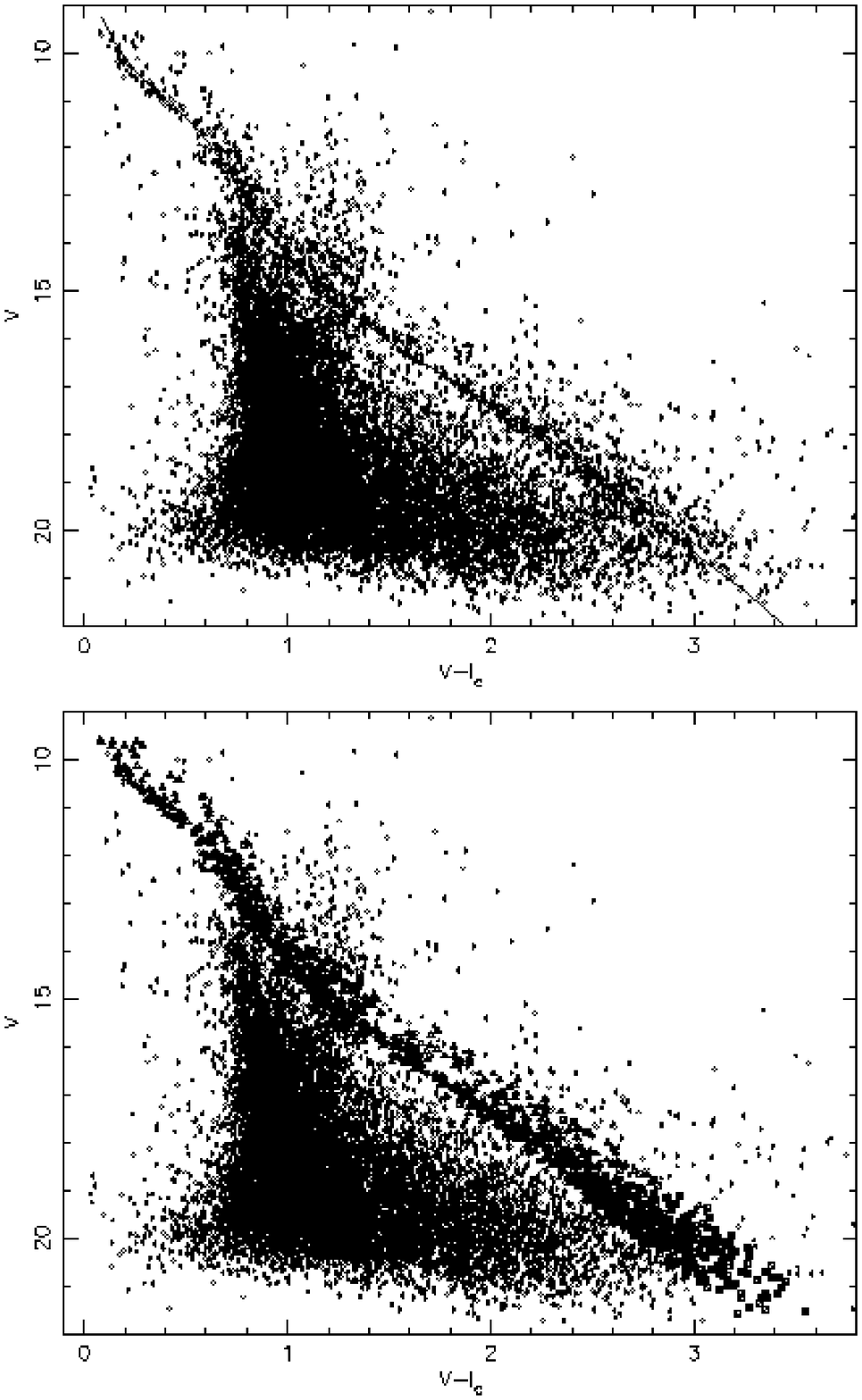}
\caption{Top: The $V$ vs $V-I_\mathrm{c}$ CMD for our survey. The solid line shows
the \citet{siess00} solar metallicity 
isochrone we use to select members of the cluster (see
Sect.~\ref{isochrone}). Bottom: Objects selected as cluster members
(see Sect.~\ref{membership})
are shown as triangles (where both $B-V$ and $V-I_\mathrm{c}$ were
available) or squares (where only $V-I_\mathrm{c}$ was available -- for 
368 objects).}
\label{vi}
\end{figure*}
\begin{figure*}
\vspace*{22cm}
\includegraphics{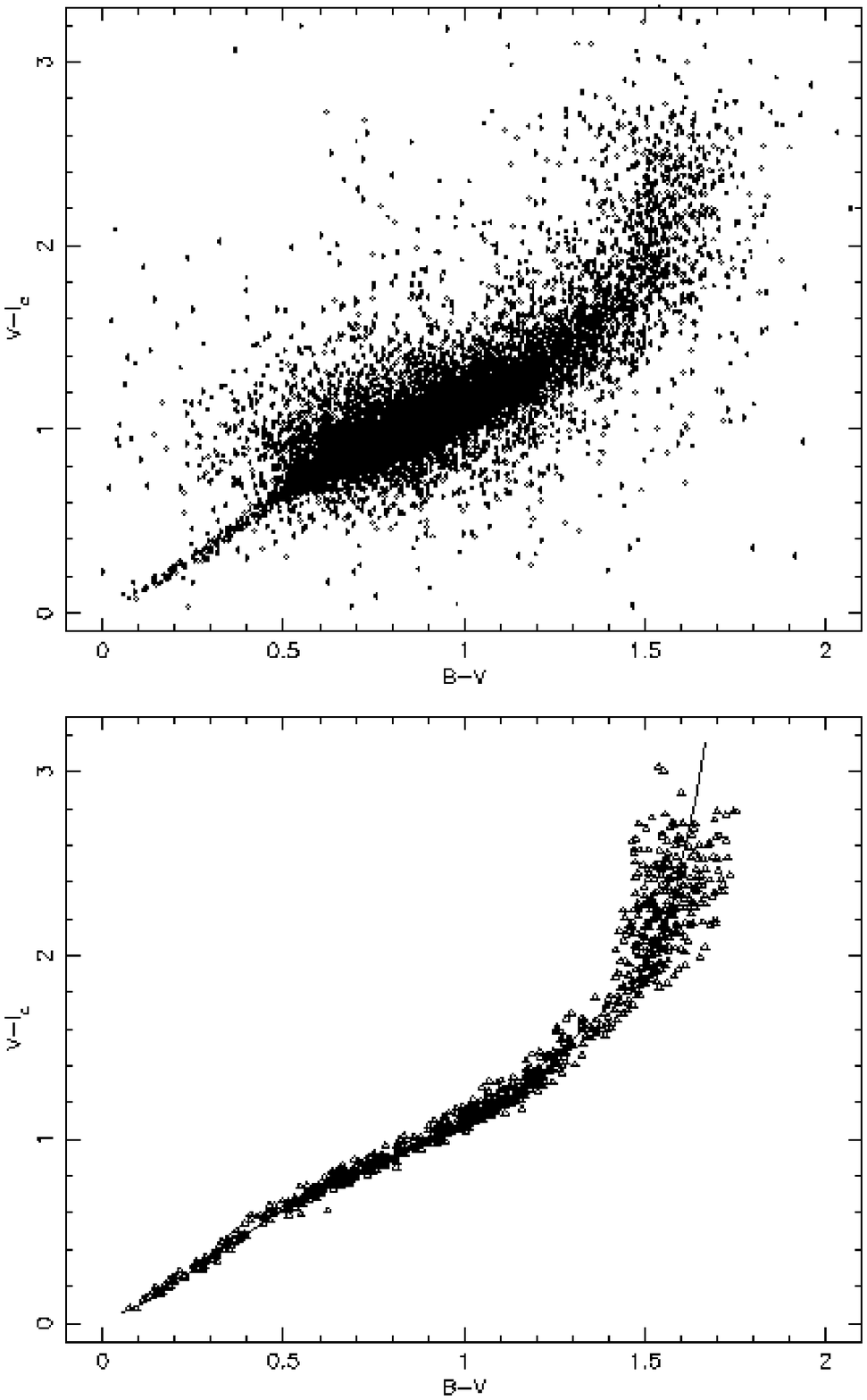}
\caption{Top: The $V-I_\mathrm{c}$ vs $B-V$ colour-colour diagram for 
our survey. Bottom: The 881 objects selected as cluster members
and which have both $B-V$ and $V-I_\mathrm{c}$ measurements 
are shown as triangles. The solid line shows
the \citet{siess00} solar metallicity isochrone we use to select these members (see
Sect.~\ref{membership}).}
\label{bvvi}
\end{figure*}

The Starlink {\sc photom} package \citep{eaton00}
was used to determine aperture
photometry of the standards in a 6 arcsec radius (the stellar FWHM for all our
data was around 1.3-1.7 arcseconds), which contained about 95-98 percent of
the stellar flux.  Nightly transformation coefficients, extinction and
zero points were determined using iterative weighted least squares fits
to photometry of about 100 \citet{landolt92} standards, rejecting stars
greater than $3\sigma$ from the fit at each step until the solution
converged.

Equations of the form
\begin{eqnarray}
V & = & v + \phi_{V}(B-V) - K_{V}X + Z_{V}\, , \label{veq}\\
B-V & = & \phi_{BV}(b-v) - K_{BV}X + Z_{BV}\, , \label{bveq}\\
V-I_\mathrm{c} & = & \phi_{VI}(v-i) - K_{VI}X + Z_{VI}\, , \label{vieq}
\end{eqnarray}
were used to transform the instrumental magnitudes (in small letters) onto the Johnson
$V$, $B-V$, Cousins $V-I_{c}$ system, where $\phi$ are the colour
coefficients, $K$ the extinction coefficients, $X$ the airmass and $Z$
the zero points (the magnitude yielding 1 detected photon per second). 
The best fit values for these coefficients on each
night are given in Table~\ref{tab1}, along with the number of Landolt standards 
measured and the rms discrepancy between their
measured and published magnitudes. Residuals were examined as a function
of airmass, colour and time with no significant trends seen.
On each night we observed several red standards
with $2<V-I_{c}<2.8$, so we expect our calibration to be well
constrained in this region.

\begin{table}
\caption{Nightly photometric coefficients, with number of standards and
root mean square discrepancies}
\begin{tabular}{lccccc}
&&&&&\\
$V$ eqn&$\phi_V$&$K_V$&$Z_V$& $N$ rms\\
8 Jan & 0.018 & 0.158 & 23.337 & 88 &0.015\\
9 Jan & 0.011 & 0.152 & 23.327 & 101&0.016\\
12 Jan& 0.022 & 0.227 & 23.277 & 111&0.024\\
&&&&&\\
$B-V$ eqn&$\phi_{BV}$& $K_{BV}$&$Z_{BV}$& $N$ & rms \\
8 Jan & 0.928 & 0.106 & -0.205 & 87 & 0.017\\
9 Jan & 0.933 & 0.120 & -0.206 &102 & 0.019\\
12 Jan& 0.934 & 0.146 & -0.205 &110 & 0.017\\
&&&&&\\
$V-I$ eqn&$\phi_{VI}$&$K_{VI}$&$Z_{VI}$& $N$ & rms \\
8 Jan & 1.001 & 0.091 & 0.968 & 86 & 0.018\\
9 Jan & 1.003 & 0.130 & 0.965 &100 & 0.024\\
12 Jan& 1.016 & 0.117 & 0.968 & 99 & 0.018\\
\end{tabular}
\label{tab1}
\end{table}

Each target frame was then analysed. The first step was to mask out
regions that were saturated due to the presence of very bright stars.
The routine {\sc daofind} from the {\sc daophot ii} package
\citep{stetson87, eaton96} was then
executed using a 5 sigma threshold and with shape rejection parameters
set to exclude extended objects and cosmic ray spikes. The number of
sources found per field ranged from 100-200 in the short $B$
exposures to about 3000 in the long $I$ exposures.

Aperture photometry was performed using the {\sc photom} routines. Both
3 and 6 arcsec radius apertures were used. For the brighter stars, where
the statistical errors 	were smaller than 0.01 mag (about $V=17$),
the larger aperture results were used and these also calibrate the aperture
correction from the small to the larger aperture for the fainter stars.
The sky estimation was taken from the mode in a surrounding annulus.
At this stage stars were rejected if masked pixels fell within the
object aperture and if they were close to the frame edges.

The next stage combined the separate filter measurements in each field,
treating the short and long exposure sequences separately.  The
instrumental $b,v,i$ measurements for a particular field were combined
by matching the pixel coordinates and allowing for small translations
between the fields. The $v$ object list was used as the reference frame
(i.e. an object must be detected in $v$ and at least one other filter 
to be incorporated in the final
catalogue).  Mean airmasses were calculated for the $B,V$ and $V,I$
pairs and the catalogue of measurements for that field transformed onto
the standard magnitude system. If no $B-V$ measurement  was available (faint, red
objects), then we used a mean relationship between $B-V$
and $V-I_\mathrm{c}$ in order to estimate $B-V$ for the purposes of
transformation to a standard $V$ magnitude (see Sect.~\ref{membership}). 
This latter procedure
should add no more than $\pm0.01$ magnitudes to the $V$ error because
the colour term, $\phi_{V}$, is small.

A preliminary astrometric solution was found for the centroided
CCD positions in the long $V$ exposure of each field. This was obtained
by identifying many stars from the Guide Star Catalogue Version 1.1
\citep{hstgscref}. A 6-coefficient fit to these reference stars,
resulted in solutions good at the level of 1 arcsecond. The x,y pixel
coordinates of objects identified in the long exposures 
were then transformed to RA and Dec using this solution. Of course some
bright stars ($V<14$) were saturated in the long exposures and their positions
must be taken from the short $V$ exposure. This was achieved by
fitting a linear translation between the long and short exposures, using
stars detected in both frames.

Objects were then matched between the catalogues from the short and
long exposures. Where objects appear in both, a weighted mean of
the photometry was
taken. There were generally about 200 well measured stars common to
both the short and long catalogues in each field. The magnitudes of
these stars were in very close agreement. The biggest differences we
ever found were about 0.02 mag, which confirmed our preliminary
assessment at the telescope that these were very good photometric
nights.

We followed a similar procedure to deal with stars in the
overlapping regions between fields. Again, where two (or even more)
measurements existed, a weighted mean of the photometry was taken. Analysis of the
discrepancies between magnitudes of stars in these overlap regions is
our primary estimate of the internal accuracy of our photometry. We
found no evidence for systematic variation in the photometric
calibrations between nights or between fields at a level greater than
0.02 mag. Table~\ref{errors} shows the internal error estimates (the rms values 
for the overlap discrepancies) as
a function of magnitude. Beyond $V=20$ the statistical errors in the
photometry rise rapidly. 

Possible causes of error are (of course) the statistical errors, but
also variations in the point spread function, and hence the aperture correction,
over the CCD field of view (especially in frames not precisely in
focus). This would contribute a term (in quadrature) which was
the same at all magnitudes -- which is approximately what we see.
Although the short and long exposures in each field were taken in one
observing sequence, some of the overlapping fields are separated by
hours or even nights. It is therefore quite plausible that some genuine
variability also contributes to these errors. That the errors
in the $V$ magnitudes are larger than the colour indices, suggests that
errors in correcting measured magnitudes to the standard star aperture
values near the edges of the CCD fields are the more likely culprit.
If this is the case, the errors in Table~\ref{errors} are likely to be
overestimates for the majority of stars with $V<18$. For $V>18$, the
statistical errors dominate.

\begin{table}
\caption{An estimate of photometric errors from stars measured more
than once in different frames}
\begin{tabular}{ccccccc}
 & \multicolumn{2}{c}{$V$ data} & \multicolumn{2}{c}{$B-V$ data} &
    \multicolumn{2}{c}{$V-I$ data} \\
   $V$ &   rms &  N& rms &  N& rms &  N\\
 11-12 &   0.078 &   15 &  0.025 &   12 &  0.036 &   13\\
 12-13 &   0.021 &   25 &  0.012 &   25 &  0.036 &   25\\
 13-14 &   0.042 &   62 &  0.013 &   58 &  0.027 &   60\\
 14-15 &   0.029 &   80 &  0.015 &   78 &  0.016 &   80\\
 15-16 &   0.032 &  163 &  0.021 &  156 &  0.033 &  156\\
 16-17 &   0.040 &  253 &  0.026 &  241 &  0.027 &  249\\
 17-18 &   0.050 &  395 &  0.041 &  318 &  0.042 &  360\\
 18-19 &   0.075 &  598 &  0.046 &  257 &  0.050 &  401\\
 19-20 &   0.099 &  534 &  0.052 &   65 &  0.055 &  255\\
\end{tabular}
\label{errors}
\end{table}

\subsection{Astrometry}
\label{astrom}

The preliminary positions determined from 6--coefficient fits to
objects in the Guide Star Catalogue (rms typically $\sim1$~arcsec) were
improved upon using pre--release data from the SuperCOSMOS Sky Survey
\citep{hambly01a}. The UK Schmidt $B_\mathrm{J}$ plate from field 124
was used (plate number J2978, epoch 1977.223). These data consisted of
a catalogue of $\sim1.4$ million objects to $B_\mathrm{J}\sim22$. At
the time these data were used, the global astrometric plate reductions
were based on standards from the Tycho--AC catalogue \citep*{urban98}
and had typical residuals per standard of $\sim0.3$~arcsec in either
co-ordinate (for more details of the astrometric reductions for
SuperCOSMOS Sky Survey data, see \citealp{hambly01b}).  For each CCD
frame, objects in the photometric catalogue were matched with objects
found on the photographic plate. A matching radius of 5~arcsec was
used. A 6--coefficient linear transformation was then applied to the
CCD coordinates with an additional cubic radial distortion coefficient
as a further free parameter. The optical axis was assumed to be the
centre of the CCD. Parameters were adjusted to get the smallest rms
when compared with tangent--plane positions on the photographic plate
(the radial distortion term was almost negligible).  The typical
zero-point shifts applied to the preliminary positions were about
0.7~arcsec, with final rms values of around 0.3 arcsecs, which we
expect are largely dominated by uncertainties in the photographic
positions.

The quality of the astrometry has recently been tested with a
fibre-spectroscopy run on the low-mass cluster candidates (see
Sect.~\ref{membership}), using {\sc hydra} on the CTIO 4-m
telescope. Excellent results were achieved over a 40 arcminute diameter
field, using brighter cluster members as the fiducial acquisition stars
and 2 arcsecond diameter fibres (Jeffries et al. in preparation).

\subsection{The catalogue completeness}

\label{complete}

The photometric/astrometric catalogue is given in Table~\ref{catalogue}
(available from the Centre de Donn\'ees astronomiques de Strasbourg) and consists of
15\,495 stars with $V$ magnitudes, $B-V$ and $V-I_\mathrm{c}$ colours when
available, along with their J2000.0 positions and flags indicating
their membership and binarity status (see Sect.~\ref{membership}). 
The complete $V$ vs $B-V$ and $V$ vs $V-I_\mathrm{
c}$ colour-magnitude diagrams (the $BV$ and $VI$ CMDs) are shown in Figs.~\ref{bv} and
\ref{vi} and the colour-colour diagram is shown in Fig.~\ref{bvvi}.
We refer hereafter to this as the CTIO catalogue.

\begin{table}
\caption{The photometric catalogue for NGC 2516, including J2000.0
positions, $V$, $B-V$ and $V-I_\mathrm{c}$ photometry where available,
flags indicating whether each star passes the three membership tests
discussed in Sect.~\ref{membership}, a flag indicating whether the star
is a candidate cluster member and finally a flag indicating whether a
cluster candidate is a binary system. The Table is available in
electronic format by ftp from CDS at cdsarc.u-strasbg.fr.
}
\label{catalogue}
\end{table}

\begin{figure}
\vspace*{8.5cm}
\includegraphics{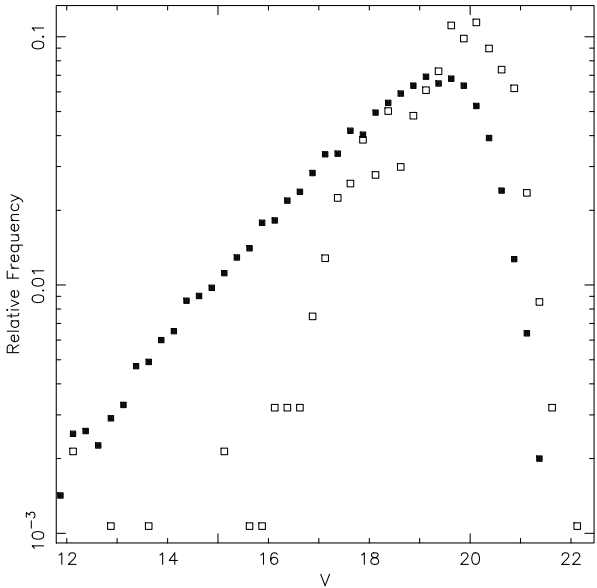}
\caption{The apparent $V$ luminosity function for all the stars
detected in our survey (solid squares) compared with the luminosity
function for those stars with $V-I_\mathrm{c}>2.0$.
}
\label{completeness}
\end{figure}

In Fig.~\ref{completeness}, the apparent $V$ magnitude luminosity function of
the whole catalogue is shown in 0.25 magnitude bins. 
There is a clear turnover in the apparent luminosity
function beyond $V=19.25$. Catalogue completeness has been investigated
by adding simulated stars to our observed frames and then searching for
these stars using the same {\sc daofind} parameters. We find that there
are three possible causes for the apparent turnover in the luminosity
function.
\begin{enumerate}
\item Clearly our search threshold imposes a magnitude limit on the
catalogue. Tests on the long $V$ frames near the centre of the cluster
reveal that the level of incompleteness caused by this and also by
faint stars being close to bright stars is approximately 1.5 percent at
$V=19$ and 10.7 percent at $V=20$. This is not sufficient to account
for the observed turnover.
\item We require that in addition to being detected in the
$V$ observation, a star is also detected in at least one of the $B$ or
$I$ observations. The individual 5-sigma detection limits for our long
exposures average about $V=20.5$, $I_\mathrm{c}=19.3$ and $B=20.4$
(depending on seeing and moon phase). Thus
in practice their are only a few (faint blue) stars detected in $V$
and $B$ only and the vast majority of the catalogue has a $V-I_\mathrm{c}$
measurement. However, there will be a significant number of
faint, intermediate colour stars ($0.5<V-I_\mathrm{c}<1.5$) which are
detected in $V$ but not $I$. Our simulations show that for $V-I_\mathrm{
c}=0.9$ (where the bulk of the population might be expected to lie),
10 and 50 percent incompleteness occur at about $I=18.9$ and $I=19.2$ and therefore
$V=19.8$ and $V=20.1$.
\item This latter effect is confused by the probability that the
turnover in the luminosity function is real. At $V-I_\mathrm{c}=0.9$, the
major population of the CMD at $V=20$ will be K-dwarfs, because the
cluster has a galactic latitude of -15.8 degrees. At $V=20$ a typical
K-dwarf would be at a distance of 7.5\,kpc and
therefore more than 2\,kpc below the galactic plane. The scale height
of such stars is actually less than 1\,kpc, so we would naturally
expect to see a decrease in the luminosity function beyond
$V\simeq18.8$. 
\end{enumerate}

The latter two effects become much weaker for redder stars. These are
invariably detected at $I$ and because M-dwarfs are much less luminous
than K-dwarfs, the scale height argument is also less applicable.
The completeness limit for faint, red stars in NGC 2516 will therefore
be determined solely by the first factor. This is illustrated in
Fig.~\ref{completeness}, where we also show the apparent $V$ luminosity
function for those stars with $V-I_\mathrm{c}>2.0$.
The final outcome of this discussion is that we believe our catalogue is almost ($>98$
percent) complete to $V=19$ over the whole $V$ vs $V-I_\mathrm{c}$
CMD. However, we estimate that our catalogue is still 90 percent complete to $V=20$ for
faint members of NGC 2516 that are selected on the basis of $V$ and
$V-I_\mathrm{c}$. If we require selection in the $V$ vs $B-V$ CMD as well,
the completeness level is governed by the $B$ frame completeness for
red stars and we have 90 percent completeness for faint, red NGC 2516 members at
about $V=18.2$, and about $V=19$ for the general field population at $B-V=0.8$.

\subsection{Comparison with other photometry}

\begin{figure*}
\vspace*{14cm}
\includegraphics{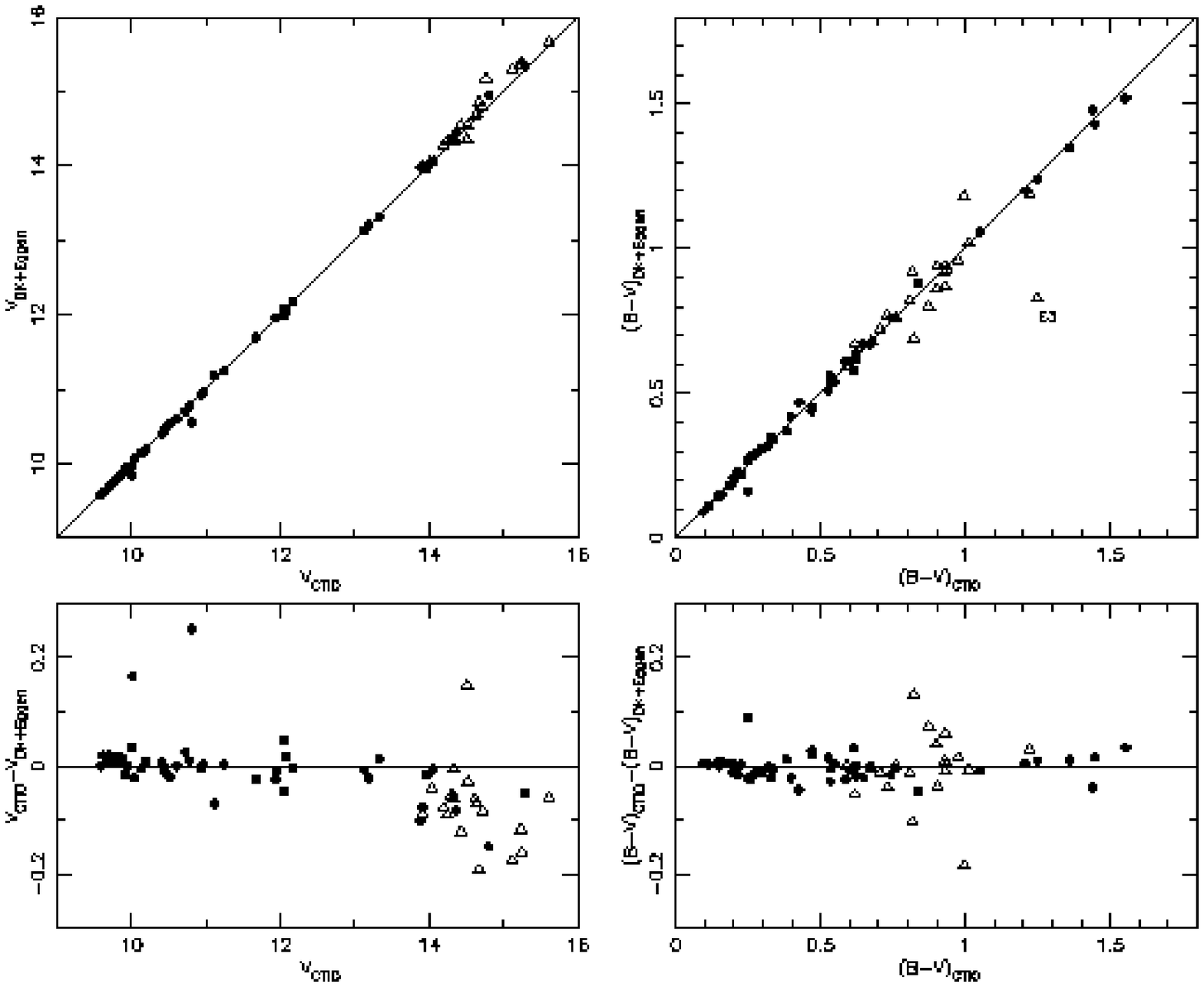}
\caption{
A comparison of photometry between this paper and photoelectric
photometry from \citet{dachs89} [solid symbols] and \citet{eggen72}
[open symbols]. The straight line simply indicates equality and is not a
fit to the data.}
\label{dkpecomp}
\end{figure*}
\begin{figure*}
\vspace*{14cm}
\includegraphics{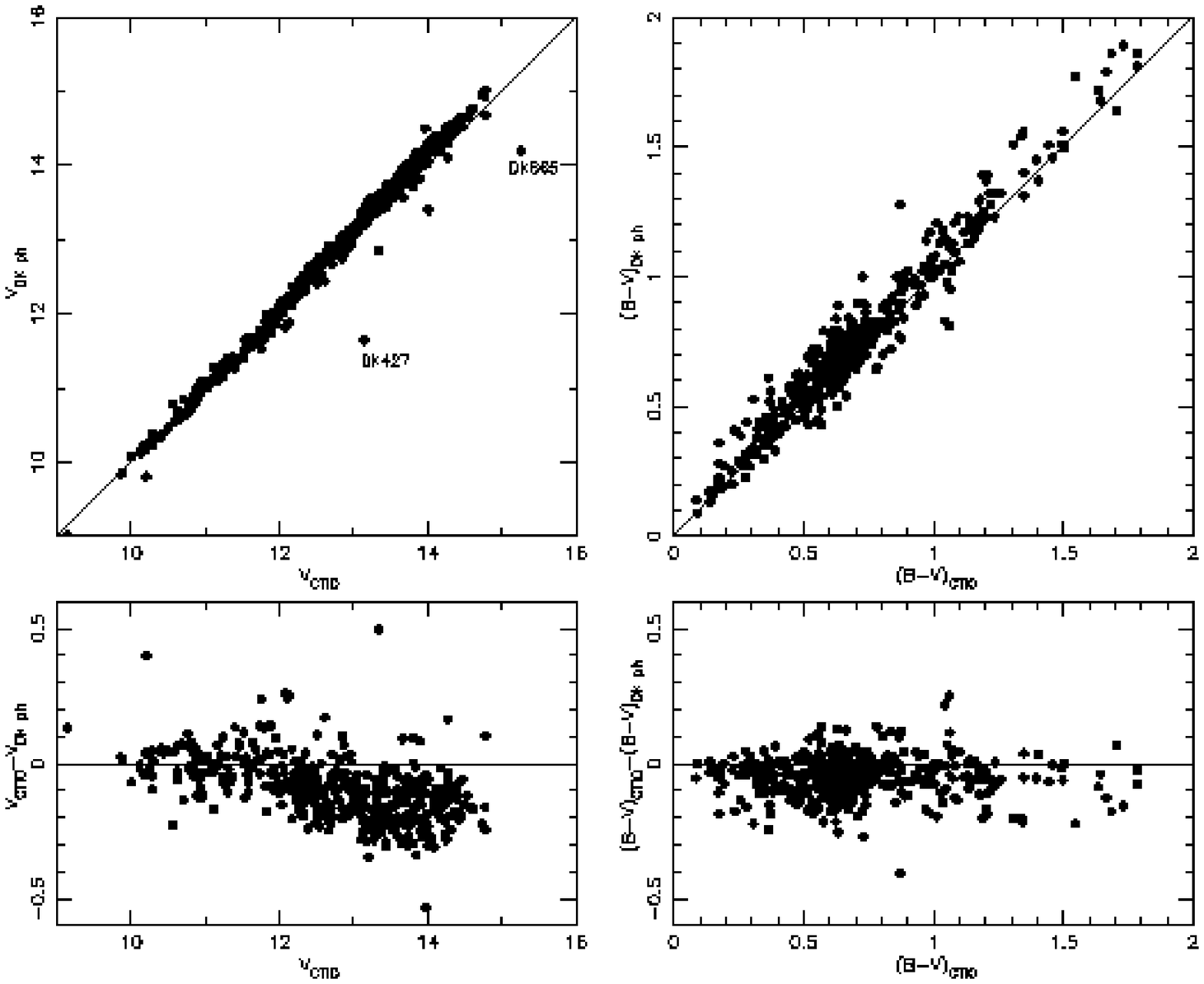}
\caption{
A comparison of photometry between this paper and photographic
photometry from \citet{dachs89}.
The straight line simply indicates equality and is not a
fit to the data.}
\label{dkphcomp}
\end{figure*}
\begin{figure*}
\vspace*{14cm}
\includegraphics{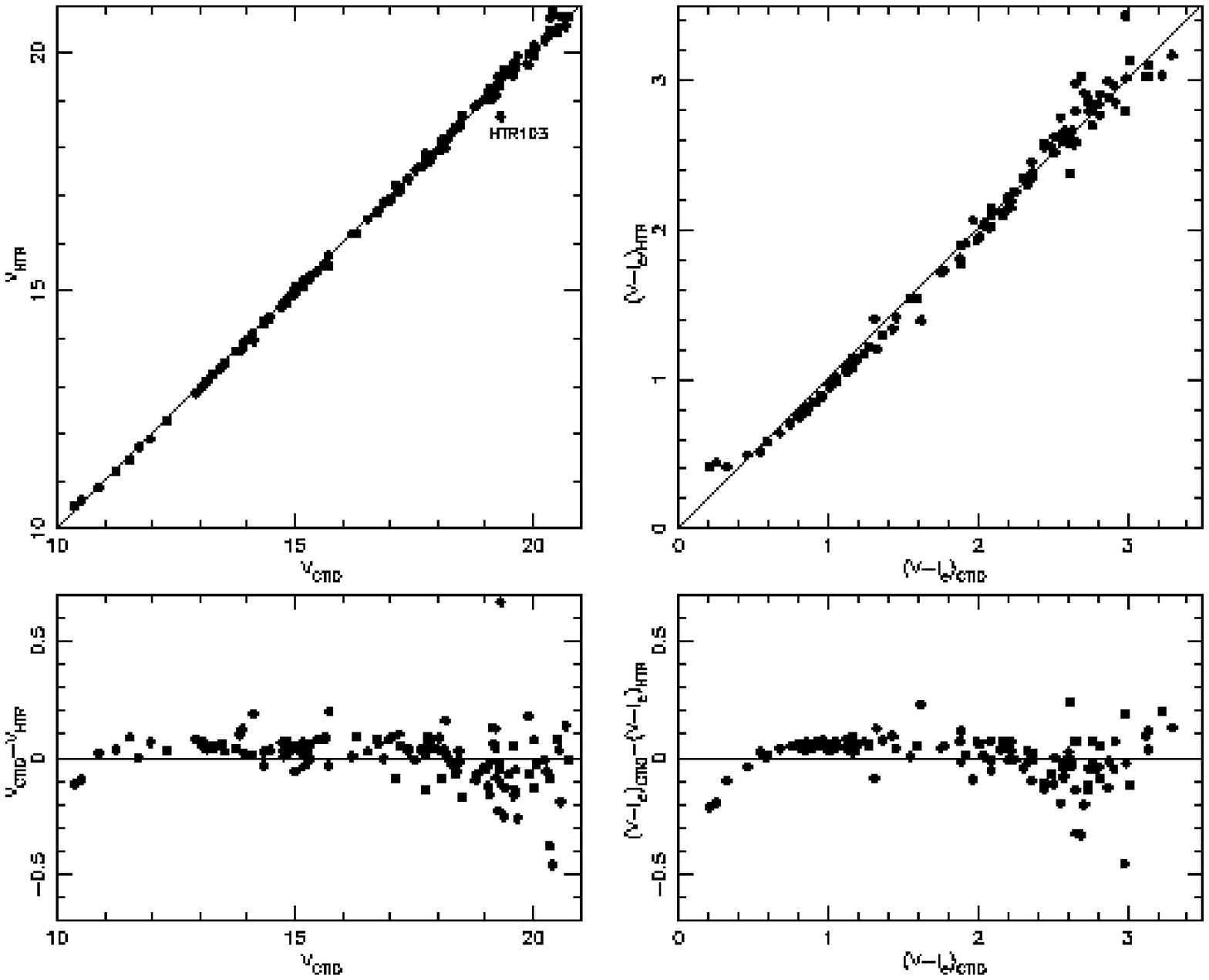}
\caption{
A comparison of photometry between this paper and CCD photometry
from \citet{hawley99}. The straight line simply indicates equality and is not a
fit to the data.}
\label{hawleycomp}
\end{figure*}

A number of other photometric studies have been published on NGC 2516,
though none that go both as deep and cover as large an area as this
one. We have compared our work with photoelectric ($BV$) photometry published
by \citet{eggen72}, photographic and photoelectric ($BV$) photometry published by
\citet{dachs89} and CCD ($VI_\mathrm{c}$) photometry published by
\citet{hawley99}.

The \citet{eggen72} and \citet{dachs89} $BV$ photometry is discussed
extensively by \citet{jeffriesn251697}. In that paper an optical
catalogue is compiled from these sources, in which small corrections (of
order 0.02 mag.) are made to the Eggen photoelectric and the Dachs \&
Kabus photographic values, to put them on a consistent system defined
by the Dachs \& Kabus photoelectric values. Dachs \& Kabus estimate
0.02 mag. errors in their photoelectric values and Jeffries et
al. deduced errors of about 0.06
mag. for their photographic values. The final catalogue
contains 568 objects and was estimated to be complete to about $V=13.5$.
We find that there are 500 matches with our new CTIO catalogue. The
remaining objects either lie outside the area covered by the CTIO survey
or were bright objects and hence saturated. \citet{hawley99}
present a list of 155 candidate members of NGC 2516 that are within our
survey area. We were able to find 130 matches to these within 1.2
arcseconds. The remaining
unmatched stars are very faint and were unlikely to be detected in our survey.

Graphical comparisons between the catalogues are shown in
Figs.~\ref{dkpecomp}, \ref{dkphcomp} and \ref{hawleycomp} 
where we have divided the comparison between: the
photoelectric data from \citet{eggen72} and \citet{dachs89} (which
are given different symbols); the photographic photometry from Dachs
\& Kabus; and the CCD photometry from \citet{hawley99}. 
The lower panels in these Figs. show the discrepancies
between the CTIO and comparison photometry. In all plots, the solid
line simply represents equality between the CTIO and comparison data,
rather than a fit.

Figure~\ref{dkpecomp} shows that there is an increasing discrepancy towards
fainter $V$ magnitudes among the the \citet{dachs89} and
\citet{eggen72} 
photometry in comparison with the CTIO data. This appears to be
the case to a much severer degree in the photographic Dachs \& Kabus
data (Fig. \ref{dkphcomp}).  We are confident that our photometry is
excellently calibrated in this $V$ magnitude range, with no trends at
all to $V=16$. We note that the two labelled outliers, DK427 and DK865,
are quite close ($<6$ arcsecs) to companion stars which might easily
have made them appear brighter in the poorer quality photographic photometry. In
contrast the $B-V$ data show very good agreement with our data,
although there is some indication that the photographic colours are too
blue by about 0.05.  We have inspected our images and can see no
problems with our photometry of the star called E3 by Eggen (labelled
in Fig. \ref{dkpecomp}). This may well have been misidentified in
\citet{eggen72}, 
which is reinforced by the agreement of our value with that
obtained by \citet{hawley99} for the same star. 
The scatter in the residuals for Figs. \ref{dkpecomp} and
\ref{dkphcomp} are in line with our estimates of the errors in these
datasets, although the Eggen photoelectric photometry appears to show
more scatter than that of Dachs \& Kabus.

Turning to Fig. \ref{hawleycomp} we again have reasonable
agreement. There is a systematic offset of about 0.04 magnitudes in $V$
(in the sense that the Hawley et al. photometry is brighter) and a
suggestion that this reverses for the faintest stars in the sample.  We
note that as Hawley et al. only published data for cluster candidates,
all these very faint stars are also very red. It therefore suggests
that there may be minor problems with either their or our
colour-dependent terms in the transformation equations. We re-iterate
that we \emph{did} observe \citet{landolt92} stars as red as $V-I_\mathrm{
c}=2.8$ and we are certainly confident in our calibration to this point
(and the colour term is very small in any case for our CCD and filter
combination).  This pattern is repeated for the $V-I_\mathrm{c}$
comparison, where Hawley et al.'s colours are blue by about 0.04
magnitudes compared with ours, with a definite reversal in the
brightest stars. The scatter in the residuals of both comparisons is
almost precisely in accord with the error estimates given by Hawley et
al. and in this paper. We have no explanation for the very discrepant
point HTR103 (labelled on the diagram), other than perhaps a major
stellar flare occurring during Hawley et al.'s observation.

\section{Selection of Candidate Cluster Members}

\subsection{Selection philosophy}

We have used the photometric catalogue to attempt a preliminary
selection of cluster members based \emph{only} on photometric
criteria. This selection procedure is especially useful because it does
not rely on any characteristic of the cluster members which one might
choose to investigate -- for instance coronal X-ray emission. Such a catalogue
of cluster candidates will be unbiased with respect to
magnetic activity, rotation rate or lithium depletion and therefore
provides an ideal starting point for such investigations
\citep[see][]{jeffriesn251698, micela00, harnden01, sciortino01}.

We must provide a caveat here; our selection
procedure is arbitrary to some degree, and \emph{will} exclude some
genuine members and \emph{will} include some non-members. We will
fashion our photometric selection criteria so as to avoid excluding the
vast majority of cluster members. The interested reader should easily
be able to generate membership catalogues using their own
(possibly more restrictive) criteria. Ideally these catalogues should
then be refined using other unbiased indicators of membership such as
proper-motions or radial velocities.

Our CCD data combined with the SuperCOSMOS scan of the 1977 epoch
Schmidt plate (see Sect.~\ref{astrom}) did allow a preliminary
attempt at proper-motion selection for the fraction of the catalogue
which had good photographic positions. We found that the best
accuracies achievable were about 6 milli-arcseconds\,year$^{-1}$ in each coordinate. We
calibrated the cluster mean proper motion using known members from
\citet{jeffriesn251698}, and found it to be essentially zero within the
errors. It was soon discovered that even for the most
accurate data, we were unable to exclude more than about 10 percent of
the general field background contamination whilst including more than
90 percent of the cluster members. We will not present these
preliminary proper motion results in this paper, but will await a
second epoch CCD survey which should be capable of producing more
precise and useful results.

Our selection philosophy is therefore restricted to using the two CMDs
($V$ vs $B-V$ and $V$ vs $V-I_\mathrm{c}$) to select stars close to a
cluster isochrone and then to check the colour-colour diagram for
consistency. 

\subsection{Isochrone generation}
\label{isochrone}

\begin{figure}
\vspace*{17.3cm}
\includegraphics{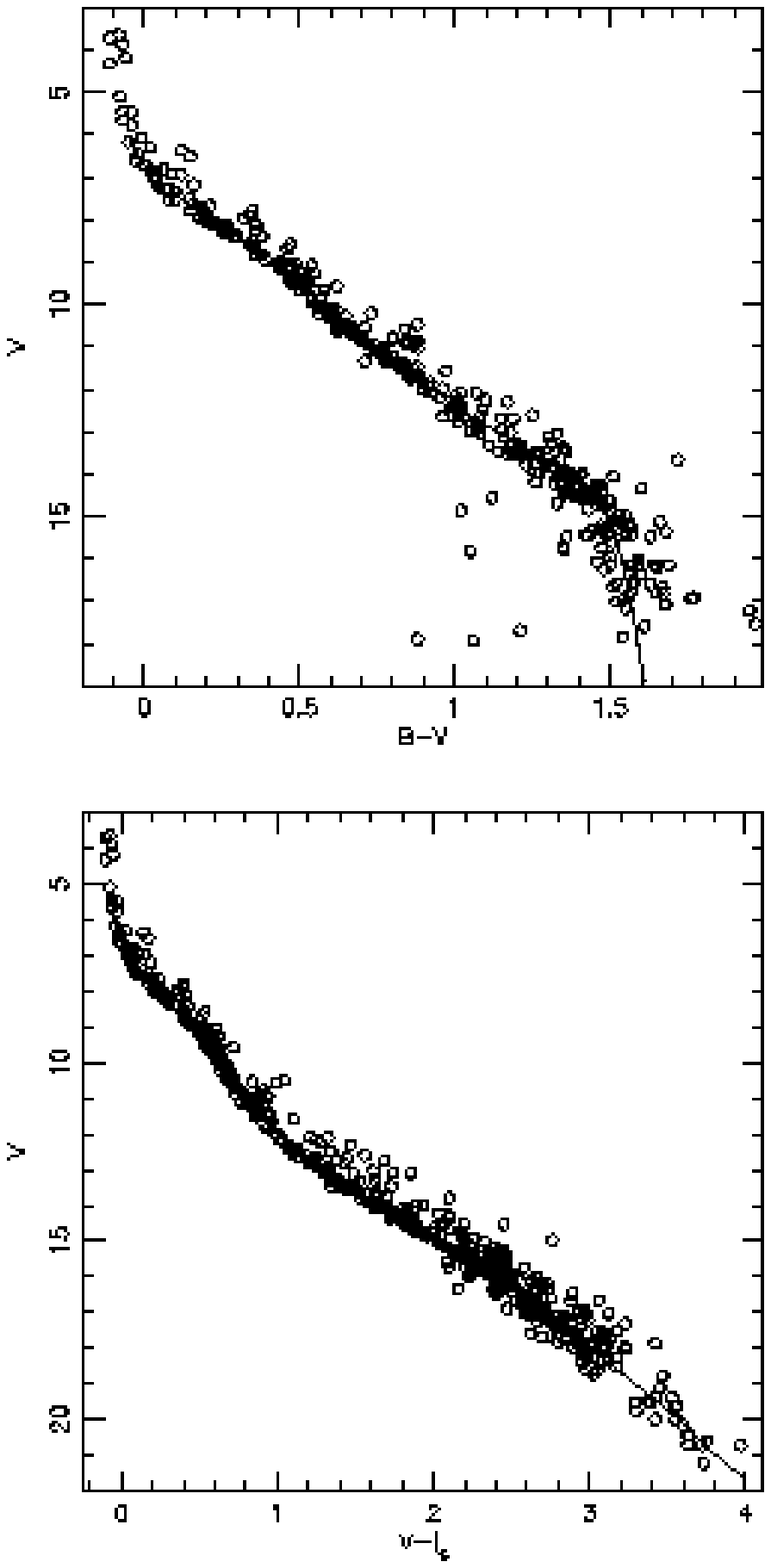}
\caption{The Pleiades $BV$ and $VI$ CMDs showing how we defined the
isochrone fits using \citet{siess00} models and a
distance modulus of 5.6 (see text).}
\label{pleiades}
\end{figure}

\begin{figure}
\vspace*{14cm}
\includegraphics{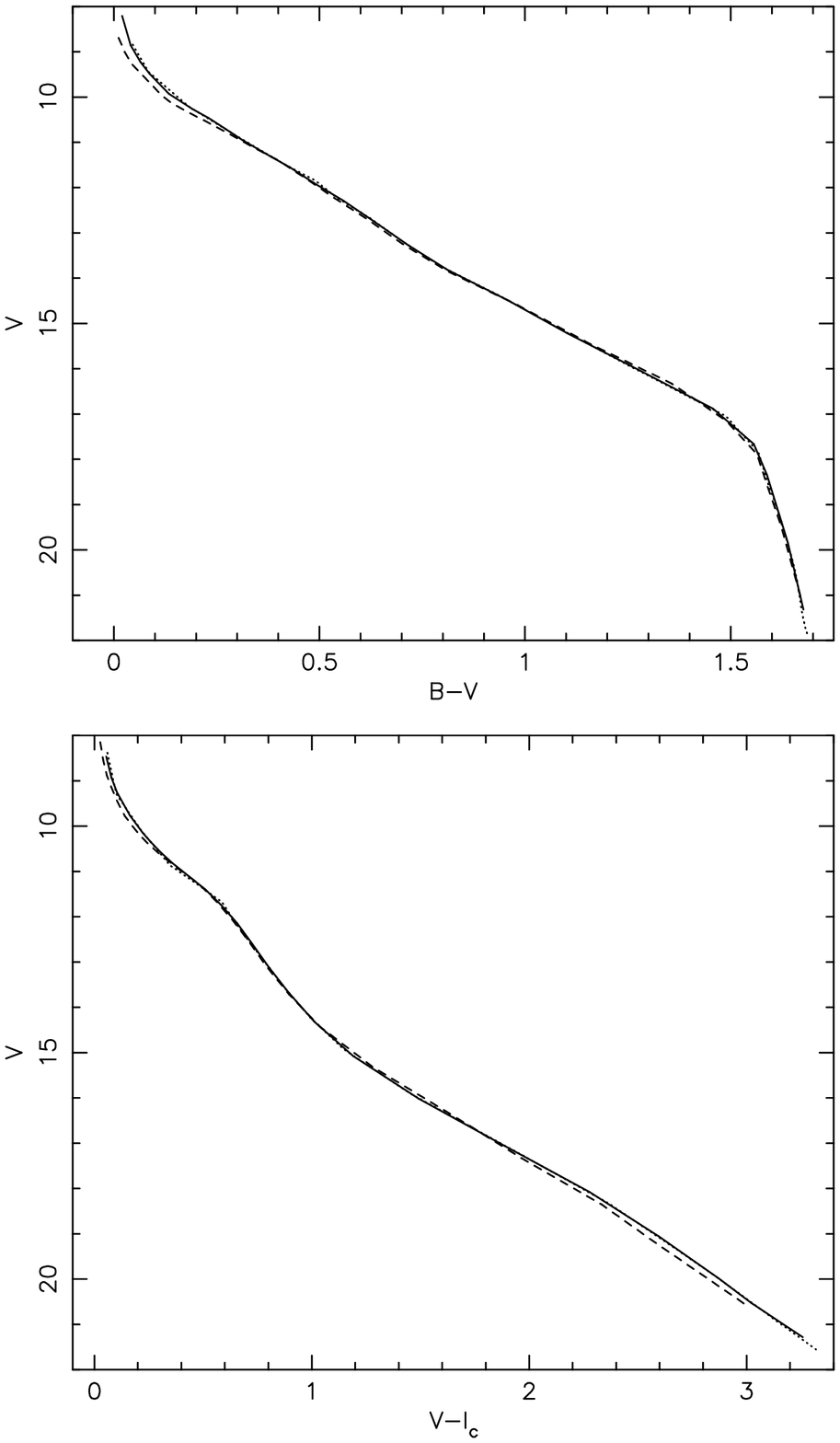}
\caption{A comparison of isochrones generated from the three models
discussed in Sect.~\ref{isochrone}. The solid lines are from the solar
metallicity model of \citet{siess00}, the dashed lines are the
half-solar metallicity models of \citet{siess00} and the dotted lines
(virtually indistinguishable from the solid line) are from the solar metallicity
models of \citet{dantona97}. The isochrones have been shifted by
different distance moduli to match the observational data (not shown
for clarity -- see Figs.~\ref{bv} and \ref{vi}).}
\label{isoplot}
\end{figure}

We used 150\,Myr approximately solar metallicity ($z=0.02$)
isochrones generated from the models of
\citet*{siess00} over the mass range
$0.1M_{\sun}<M<7M_{\sun}$. These isochrones were provided in the
form of bolometric luminosity versus effective temperature and
had to be transformed into the observational plane. We did {\em not} use
the photometric colours provided with the models.
The critical step here is the relationship between colour and effective
temperature, which is highly uncertain, especially among cooler
stars. 

The procedure we adopted is as follows: We obtained a catalogue of
$BVI$ photometry of the Pleiades (courtesy of the Open Cluster Database
operated by J. Stauffer and C. Prosser) and supplemented this with
photometry of cooler Pleiades members presented by \citet{stauffer98b}.  
$V-I$ indices on the Kron system were converted to the
Cousins values using transformations in \citet{bessell87}. We
posit that there \emph{is} a single colour-$T_\mathrm{eff}$ relationship
which applies to all stars of approximately the same age as the
Pleiades, for which we assume an age of 120\,Myr, a distance modulus of
5.6, a reddening of $E(B-V)=0.04$, $E(V-I_\mathrm{c})=0.05$ and extinction
of $A_{V}=0.13$ (see \citealp{stauffer98b}; \citealp*{stauffer98}).
Note that the
exact values of these quantities have very little influence on the
membership selection for NGC 2516 but could affect the deduced distance
modulus or metallicity for the cluster from isochrone fitting.

An empirical isochrone is fitted (by eye) to the Pleiades
data (in both $BV$ and $VI$ CMDs), ensuring that the isochrone is not
biased upward by the presence of binary cluster members (see
Fig.~\ref{pleiades}).   
At about 25 fiducial points along these empirical isochrones
we determine absolute magnitudes and colours, which are then converted
to bolometric luminosities using bolometric correction-colour
relationships from polynomial fits to the empirical data in \citet{flower96}
for $B-V$, and a combination of empirical data from \citet{leggett96} for
$V-I_\mathrm{c}>0.7$ and the atmospheric models of \citet*{bessell98} 
for $V-I_\mathrm{c}<0.7$. Judging by the scatter around these
relationships and by comparisons with other bolometric correction-colour
relationships \citep[e.g.][]{monet92} we estimate systematic errors 
of no more than a few hundredths of a magnitude (at
least over the colour range we are interested in). This contrasts
markedly with the uncertainties if one tries to use a bolometric
correction-$T_\mathrm{eff}$ relation, which is often what has been done in
the literature to convert models to observables.

Having obtained luminosity as a function of intrinsic colour, we then
use the assumed age of the Pleiades to obtain the $T_\mathrm{eff}$ value
appropriate for a particular luminosity, by interpolating along the
120\,Myr model isochrone. This in turn defines a
set of colour-$T_\mathrm{eff}$ points which can then be used to transform
any other model isochrone.

The net result is that we have well calibrated isochrones over the
colour range defined by the fiducial Pleiades data (or the mass range
of the models -- whichever is more restrictive). In practice this
means $-0.15<(B-V)_{0}<1.7$ and $0.4<(V-I_\mathrm{c})_{0}<4.0$. The hotter
limit to the $V-I_\mathrm{c}$ range is defined by a lack of $V-I_\mathrm{c}$
photometry for hotter Pleiades stars. For the sake of defining an
isochrone for membership selection, we have extended this data back to
$(V-I_\mathrm{c})_{0}=-0.1$ by transforming the hot star $B-V$ data into
$V-I_\mathrm{c}$ using relationships defined by \citet{johnson66}
(between $B-V$ and Johnson $V-I$) and then \citet{bessell79} (between
Johnson $V-I$ and $V-I_\mathrm{c})$.

The 150\,Myr isochrones were transformed into the $BV$ and $VI$
CMDs, and reddening and extinction applied for NGC 2516 ($A_{V}=0.38$,
$E(B-V)=0.12$ -- see \citet{dachs89} and \citet{jeffriesn251697, jeffriesn251698}
-- and an assumed $E(V-I)=0.15$). We then adjusted the intrinsic distance
modulus of the isochrones to match the data, particularly for $B-V<0.9$
in the $BV$ CMD and for $V-I_\mathrm{c}<0.8$ and $1.4<V-I_\mathrm{c}<2.2$
in the $VI$ CMD, where contamination by field stars appears to be
small (see Sect. 3.4). The results are shown in Figs. \ref{bv} and \ref{vi}, where a
distance modulus of 8.10 has been applied to both the $BV$ and
$VI$ CMDs, with estimated uncertainties of $\pm0.05$.

This procedure is quite robust to the assumed ages of the Pleiades and
NGC 2516, because for all but the very hottest and coolest stars, changing the age
by as much as 50\,Myr makes little difference to the isochrones. It is
also robust to the choice of evolution model, because we require the
models to fit the Pleiades at a similar age. We obtain essentially
an identical fit to the NGC 2516 data using 150\,Myr isochrones from
\citet{dantona97} and with the same distance moduli as above (see Fig.~\ref{isoplot}).

Of more consequence is the possibility that the metallicity of NGC 2516 is
sub-solar by a factor of two \citep{cameron85b,
jeffriesn251698, pinsonneault00}.  If this is the case then the
distance moduli we have obtained will be overestimated because lower
metallicity stars are fainter at the same colour. The effect should be larger
in the $BV$ CMD than the $VI$ CMD because the opacity caused by
metal-lines causes blanketing in the $B$ band \citep{alonso96}. \citet{pinsonneault98}
have calibrated this effect for F and G stars and indeed
used it to calculate the metallicities (and metallicity-corrected
distance moduli) for a number of open clusters. \citet{jeffriesn251698}
and \citet{pinsonneault00} used the same approach to calculate a
metallicity for NGC 2516 of [M/H]=$-0.18\pm0.08$ and -0.26 respectively
on the basis of preliminary $BVI$ photometry of known members (from
this dataset).

We have checked the effects of a low metallicity by 
using a 150\,Myr isochrone from \citet{siess00}
with $z=0.01$ (approximately half-solar metallicity, [M/H]$\sim-0.3$). 
This is again calibrated using the
Pleiades photometry. A question arises as to whether the
colour-$T_\mathrm{eff}$ relation derived empirically from the Pleiades is suitable for
a lower metallicity cluster. The $B-V$ index {\em is} sensitive to
metallicity for warm stars with partially ionized metal lines and is
partly the reason that the $BV$ CMD may be changed more by metallicity than the $VI$
CMD. However for hot stars ($B-V<0.2$) and cool stars
($B-V>1.3$) this is not likely to be the cases
\citep[see][]{castelli99}. \citet{leggett96} also show that the
$V-I_\mathrm{c}$ colour index is a good temperature indicator with
relatively little metallicity sensitivity for cool stars
($V-I_\mathrm{c}>1.5$) and line blanketing in the $V$ band is not
expected to be very important for hotter stars. Thus to first order,
this approach to generating a low metallicity observational isochrone
should be valid.

We find that the solar and half-solar metallicity isochrones yield
comparable fits to the data.
As expected, the distance moduli required to fit the
data are smaller for the lower metallicity models. We find distance
moduli of $7.85\pm0.05$ and $7.90\pm0.05$ for the $BV$ and $VI$ CMDs
respectively. These distance moduli are in excellent
agreement with the $7.94\pm0.04$ in \citet{jeffriesn251697} and
$7.96\pm0.17$ found by \citet{jeffriesn251698}, but a little larger than
the $7.77\pm0.10$ deduced by \citet{pinsonneault00} and the Hipparcos
distance of $7.70\pm0.16$ found by \citet{robichon99}. We emphasize
that our errors are underestimated because they do not take into
account uncertainties in the metallicity or reddening.

The isochrones are compared in Fig.~\ref{isoplot}, where we also
include the solar-metallicity \citet{dantona97} model. The isochrones
have been shifted to the distance moduli required to give a reasonable fit
to the observational data (which is not shown for clarity of comparison
between the models). The most important point to make is that the
{\em shapes} of the models are extremely similar. The main discrepancy
occurs in the cool part of the $VI$ CMD, where 
low metallicity stars lie just less than 0.1 magnitudes \emph{below}
solar metallicity stars of the same colour.
This is fortunate, because it means irrespective of which
model/metallicity we choose, the selection of cluster members by
photometric means is almost unaffected -- although cluster properties
such as the mass function are (see Sect.~\ref{massfun}).
A more detailed investigation of the metallicity, distance and
reddening is left to another paper that uses a more complete sample 
of spectroscopically confirmed NGC 2516 F
and G stars \citep{terndrup01}.

\subsection{Membership criteria}
\label{membership}

To select the members we apply the following criteria, based on the
\citet{siess00} solar metallicity isochrones:
\begin{enumerate}
\item If a $V-I_\mathrm{c}$ colour is available, the star must have $V$
within a region bounded by 0.1 mag below the cluster $V$
vs $V-I_\mathrm{c}$ isochrone, and 0.85 mag (to include binary stars
and allow for the uncertain metallicity -- see Sect.~\ref{isochrone}) 
above it. Additionally we allow
an extra 2$\sigma$ error in $V$ (see Table~\ref{errors}), which also incorporates
a (usually dominant) 
contribution from the $V-I_\mathrm{c}$ error, which is added in quadrature
to the $V$ error by assuming a gradient in the CMD appropriate for a
cluster member at that colour.
\item If a $B-V$ colour is available then the star must satisfy a
similar condition in the $V$ vs $B-V$ CMD. At faint magnitudes
($V>17$), the cluster isochrone steepens. For these stars we additionally allow 
a cluster member to lie within $2\sigma_{B-V}$ of a band running
$\pm0.05$ mag in $B-V$ either side of the cluster isochrone.
\item If  $B-V$ and $V-I_\mathrm{c}$ are both available, the
star must lie within $\pm0.05$ mag. (in $V-I_\mathrm{c}$) of the cluster
locus in the colour-colour diagram (see Fig. \ref{bvvi}). Again, we
allow an extra slop of 2$\sigma$ which includes contributions from both
the $B-V$ and $V-I_\mathrm{c}$ errors added in quadrature. The $V-I_\mathrm{c}$ vs
$B-V$ cluster locus is calculated by combining and interpolating
the two CMD cluster isochrones. Redder than $V-I_{c}>1.9$ this locus
steepens and similarly to test 2 above, we allow a cluster member to lie within
$2\sigma_{B-V}$ of a band running $\pm0.05$ mag in $B-V$ either side of
the cluster locus. 
\item Finally, if the star passes test 1, it is classed as a \emph{binary}
candidate if it lies more than 0.3 mag in $V$ above the $V$ vs
$V-I_\mathrm{c}$ isochrone. This should select unresolved binaries with
mass ratios of approximately 0.6-1 (see Sect.~\ref{binary}).

\end{enumerate}

The full catalogue contains 15\,495 stars. Of these we find that 11\,114
have a $B-V$ value, 15\,310 have a $V-I_\mathrm{c}$ value and 
10\,929 have both. 1499 stars pass test 1, 1295 stars pass test 2
and 920 stars pass them both, of which 881 pass test 3.
However, in order to be classed as a cluster member
we only require that the star should not fail any of these
tests. Therefore we include 368 stars which have no $B-V$ colour
but pass test 1. These are almost all very faint, red cluster candidates.
We also include 5 stars which possess no $V-I_\mathrm{c}$ but pass test 2.
The final membership catalogue therefore contains 1254 candidates.
Figures \ref{bv}, \ref{vi} and \ref{bvvi} show these
cluster candidates with different symbols indicating how they have
been classified. Of the 1254 cluster candidates, 403 are possible
unresolved binary systems. Flags corresponding to the status of stars
with respect to these tests are appended to the catalogue in Table~\ref{catalogue}.

\subsection{Contamination by non-members}
\label{contaminate}

\begin{figure}
\vspace*{8.2cm}
\includegraphics{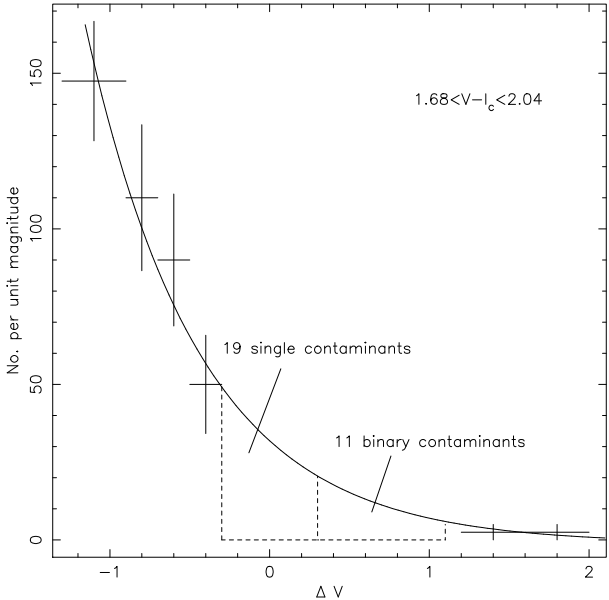}
\caption{An example of our fitting procedure to determine the level of
contamination among our candidate members (see text). The y-axis shows
the number of star per unit magnitude found in strips a distance
$\Delta V$ (on the x-axis) from the cluster isochrone. The example shown
is for strips bounded by $1.68<V-I_\mathrm{c}<2.04$ (see Table~\ref{contam}).
}
\label{contamplot}
\end{figure}

Whilst we are reasonably confident that our membership criteria should
have included the vast majority of true cluster members (within the
bounds of our photometry completeness), it is clear looking at Figs.
\ref{bv} and \ref{vi} that we have also included contaminating
foreground and background sources.
As we have taken no offset fields of the cluster, we attempt to estimate
membership probabilities and quantify this contamination by
interpolating the distribution of cluster non-members in the $VI$ CMD.

We proceed by removing the cluster candidates from the catalogue and
then select stars in several strips in the $VI$ CMD, both above and
below (but parallel to) the fiducial
cluster isochrone (from 1.2 magnitudes below to 2.0 magnitudes above), 
but beyond the limits defining membership. We 
fit a smooth function to the density (number of stars per $V$ magnitude
interval) of contaminating stars along vertical strips (colour ranges)
in the CMD.
We achieved the best looking fits using the sum of a
constant density and an exponential decay with an e-folding length of
0.5-0.9 mag. An example is shown in Fig.~\ref{contamplot}.
The fitted function is then integrated over the
range in which cluster members were selected, in
order to estimate how many contaminating stars we expect
to fall within the membership selection region. Because the
contamination tends to decrease quite sharply as  we move above the
cluster isochrone we subdivide this range further in an attempt to
estimate how much contamination there would be among single and
binary star candidates separately. This was achieved by integrating the
the interpolating function above and below a line 0.3 mag. above the
cluster isochrone (see Sect.~\ref{membership}).
These figures are then
reduced by the number of objects which passed test 1, but failed tests
2 or 3, because we argue that these must be part
of the contaminating sample. Finally we subtract the remaining
contaminants from the number of cluster candidates in the same
magnitude range and divide by the number of cluster candidates to
estimate the probability that a cluster candidate is a
genuine cluster member. At the same time we can see what
advantage has accrued from using more than one colour in assessing
cluster membership, by finding what fraction of the contamination in the
$VI$ CMD has been rejected using $B-V$. 

The results are given in Table~\ref{contam}, where we have split the sample into
colour ranges which roughly correspond to 1 magnitude $V$ intervals for
single stars in the cluster. Such crude binning is necessary in order
to have enough stars (especially in the CMD strips above the cluster
isochrone) to yield reliable fits for the interpolating function.  We
have also assumed that all hotter candidates (roughly brighter than
$V=11.5$) are members.

\begin{table*}
\caption{Estimates of contamination and membership probability among
our cluster candidates. Rows 1 and 2 list the observed numbers of
cluster candidates in each colour interval, for single stars and
binaries respectively. Rows 3 and 4 list the predicted number of
contaminants derived from our fitting technique in the $VI$ CMD (see
text). Rows 5 and 6 list how many of these contaminants are rejected
using the $B-V$ colour and rows 7 and 8 give the final membership
probabilities for candidates.}
{\small
\begin{tabular}{lccccccccc}
\multicolumn{1}{l}{$V-I_\mathrm{c}$} &0.53-0.70&0.70-0.87&0.87-1.05&1.05-1.33&1.33-1.68&1.68-2.04&2.04-2.42&2.42-2.73&2.73-3.02\\
&&&&&&&&&\\
{\bf Observed}\\
Single Candidates & 31 & 61 & 51 & 115 & 67 & 66 & 119 & 146 & 106\\
Binary Candidates & 17 & 41 & 37 &  62 & 27 & 23 &  40 &  40 &  47\\ 
&&&&&&&&&\\
{\bf Predicted}\\
Single contaminants& 6 & 38 & 40 & 63 & 14 & 19 & 25 & 63 & 31\\
Binary contaminants& 5 & 18 & 25 & 55 & 15 & 11 & 22 & 39 & 26\\
&&&&&&&&&\\
{\bf Rejected}\\
Single contaminants& 6 & 22 & 16 & 32 & 13 & 13 & 17 & 26 &  1\\
Binary contaminants& 3 &  8 &  4 & 27 & 11 & 11 &  5 & 19 & 11\\
&&&&&&&&&\\
{\bf Membership Prob.}\\
Single candidates &$100\pm11$&$74\pm13$&$53\pm16$&$73\pm9$&$98\pm8$&$91\pm9$&$93\pm5$&$75\pm7$&$72\pm6$\\
Binary candidates &$88\pm17$&$76\pm13$&$43\pm17$&$55\pm16$&$93\pm19$&$100\pm20$&$58\pm15$&$50\pm20$&$68\pm14$\\
\end{tabular}
}
\label{contam}
\end{table*}

The results in Table~\ref{contam} can now be used
to correct statistical ensembles for contamination (e.g. in investigating the
luminosity function -- see Sect.~\ref{lumfun}). We caution however that these
membership probabilities are \emph{averages} over our field of
view. Because the cluster is centrally concentrated (see Fig.~\ref{n2516opt} and
Sect.~\ref{masseg}) and we expect that the background contamination
has a uniform spatial distribution, membership probability will be higher for a
cluster candidate close to the cluster centre and lower for a candidate
near the edge of our survey area. 

The contamination is worst for $0.87<V-I_\mathrm{c}<1.33$
($12.5<V<15.5$) and a glance at Fig.~\ref{vi} confirms that this is
where the two ``fingers'' of contamination caused by main sequence and
background giant stars respectively, cross the cluster
isochrone. Candidates from $1.33<V-I_\mathrm{c}<2.42$ ($15.5<V<18.0$)
suffer very little contamination, but note that this is largely because
the addition of $B-V$ data removes the majority of the contamination in
the $VI$ CMD.  For $V-I_\mathrm{c}>2.42$ ($V>18.0$) there is growing
incompleteness in the $B$ band as well as a growing number of
contaminants from the $VI$ CMD alone, resulting in a drop in the
discrimination in our membership selection. We expect that the level of
contamination is \emph{underestimated} for the coolest bin
(approximately $V>19.5$) in Table~\ref{contam}, because the data below
the cluster isochrone in the $VI$ CMD is incomplete.  The
errors in the membership probabilities are estimated assuming Poisson
errors in the numbers of candidate members, the numbers of candidate
members rejected on the basis of their $B-V$ and the numbers of
predicted contaminants in the $VI$ CMD.  This latter error is an
overestimate, because the contaminant numbers arise from modelling
populations several times larger. In any case, these errors are similar
(as a fraction) to the simple Poissonian errors in the numbers of
cluster candidates in each colour bin.  In what follows we will simply
assume that the contamination fraction is known accurately, but will
compare our results with what would have been obtained without any
correction for contamination.

It is of interest to compare the estimates above with information
appearing in the literature. \citet{jeffriesn251698} used a similar
$BVI$ photometric selection technique to choose a cluster candidate
list. These were subsequently followed up with spectroscopy. In that
paper, 22 out of 31 objects with $12.5<V<15.0$ were confirmed as
members based on their radial velocities. This is perfectly consistent
with the estimates in Table~\ref{contam}. 

\citet{hawley99} \emph{did} do an offset field (about 1 degree from
the cluster centre) covering 225 square arcminutes in $V$ and $I$
only. It is not clear (see Sect.~\ref{masseg}) that this is far
enough from the cluster centre to guarantee no cluster members and
indeed Hawley et al. spectroscopically identified a few low mass NGC
2516 candidates in this field. However, we can use similar membership
criteria to those in our survey to find how many bogus NGC 2516 members
we might expect in a 225 square arcminute area. This is then multiplied
by 13.7 to match our survey area. The total number of contaminants
expected in the $VI$ CMD (to be compared with the sum of 
rows 3 and 4 of Table~\ref{contam})
are $192\pm51$ for $0.87<V-I_\mathrm{c}<1.68$ (compared with 212 in
Table~\ref{contam}), and $274\pm61$ for $1.68<V-I_\mathrm{c}<2.73$ (compared with 179
in Table~\ref{contam}).  The former estimate is in good agreement with our own,
but the latter is a little higher, perhaps indicating that there are
indeed some low mass NGC 2516 members even 1 degree ($\sim7$\,pc)
from the cluster centre.

\section{Luminosity and Mass Functions}

The CTIO catalogue of candidate members, together with the estimates of
contamination in Table~\ref{contam}, allow us to make a preliminary investigation
of the luminosity and mass functions (LF and MF) of NGC 2516 (that is the number of
stars per $V$ magnitude interval and the number of stars per logarithmic mass
interval).

\subsection{Luminosity Function}
\label{lumfun}

\begin{figure}
\vspace*{8.6cm}
\includegraphics{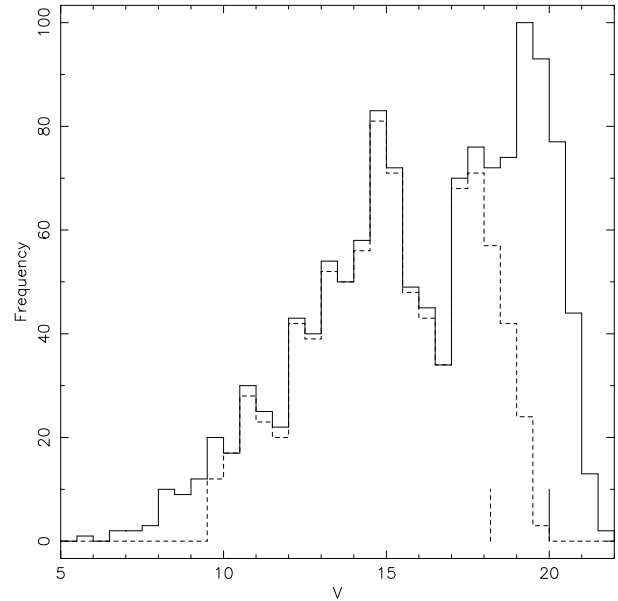}
\caption{The $V$ band luminosity function (LF) of NGC 2516. The solid histogram
shows the LF for candidate members identified from either $B-V$ or
$V-I_\mathrm{c}$ data (including brighter stars from \citealp{jeffriesn251697})
and the dashed histogram represents the LF of those stars for which both
$B-V$ and $V-I_\mathrm{c}$ indicate membership. The two short 
lines at the base of the plot 
indicate the approximate 90 percent completeness limits for these
two samples at $V=20.0$ and $V=18.2$ respectively.
}
\label{lfplot1}
\end{figure}

\begin{figure}
\vspace*{12.2cm}
\includegraphics{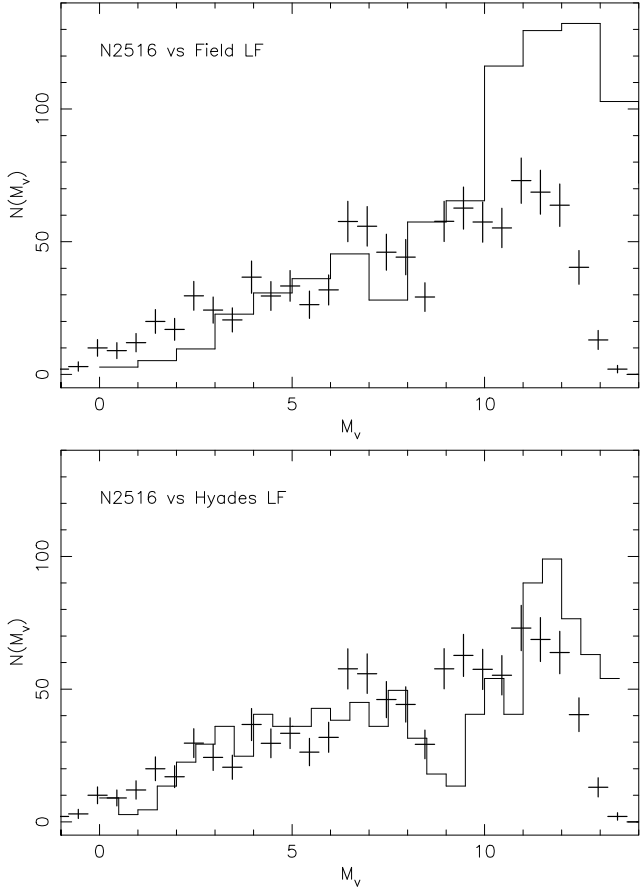}
\caption{The corrected absolute $V$ magnitude luminosity function (LF
-- see text) of NGC 2516 (shown with error bars) compared with LFs for the field and the
Hyades. The field and Hyades LFs are shown as histograms and are normalised to
match the NGC 2516 LF at $5<M_{v}<7$. The error bars on the field LF
are about the same as in NGC 2516 at the same absolute magnitude,
whereas the Hyades LF error bars are about a factor of two bigger (but
with twice as many bins). The field and Hyades LFs were taken from
\cite{reidhawley00} and references therein. 
}
\label{lfplot2}
\end{figure}

The CTIO candidate membership catalogue is incomplete for $V>20$, but
is also incomplete for brighter stars because of saturation in the CCD
frames. As discussed in Sect. 2.4, there were 68 objects from the
catalogue of bright stars in the NGC 2516 field compiled by \citet{jeffriesn251697},
which are not present in the CTIO catalogue. 55 of these objects (which
have $V$ and $B-V$ photometry) reside
within our CCD frames and 48 have photometry consistent with main
sequence cluster membership. These should be included when calculating the
LF and MF, and, because \citet{jeffriesn251697}
estimate that this bright catalogue is complete to $V=13.5$, we then
have an almost complete catalogue of cluster members from $5.8<V<20$,
covering a mass range of approximately $0.3<M<5.0M_{\sun}$ (the upper
limit being rather age, model and metallicity dependent).

Figure~\ref{lfplot1} shows the LF of the candidate
members without a correction for contamination. The plot shows the LF
due to candidates selected using \emph{both} $B-V$ and $V-I$ and also
the complete candidate list (including the brighter stars discussed
above and those stars classed as members for which either $B-V$ or
$V-I_\mathrm{c}$ are unavailable). We also mark on the plot the approximate
faint magnitude completeness limits for these two samples (discussed in
Sect. 2.3).

The rapid fall off at faint magnitudes coincides with our estimates of
where the survey begins to be significantly incomplete. There 
\emph{appears} to be a minimum in the LF for $15.5<V<17.0$, which was also
commented upon by \citet{hawley99}. However, we might suspect that
this feature merely results from the inclusion of many background
contaminants at slightly brighter magnitudes (see Table~\ref{contam}
and Fig.~\ref{vi}), resulting
in the peak at $13.5<V<15.5$. An NGC 2516 LF 
corrected for contamination by non-members and plotted with $M_{v}$ on
the x-axis (assuming an apparent distance modulus of 8.3) appears in
Fig. \ref{lfplot2}.  We compare this with LFs from the field
\citep{reidhawley00}
and the Hyades \citep{reid92, reidhawley00}, which are
normalised to match NGC 2516 for $5<M_{v}<7$.  The correction for
contamination in NGC 2516 is applied by making each star worth $\leq 1$
in the luminosity histogram according to it's $V-I_\mathrm{c}$ colour and
whether it is a photometric binary candidate, using the fraction of
such stars which are likely to be members according to Table~\ref{contam}. 
For instance, a candidate member with $V-I_\mathrm{c}=1.0$ and which is flagged as
a probable binary system will only contribute 0.43 to the LF bin
corresponding to its $V$ magnitude.  All the very bright candidate
members without $V-I_\mathrm{c}$ are assumed to be genuine members.

The NGC 2516 LF in Fig. \ref{lfplot2} exhibits both similarities and
differences when compared with the LFs of the field, the Pleiades (which is
almost identical to the field LF -- \citealp*{hambly93};
\citealp*{meusinger96}) 
and the Hyades. The LF is consistent with a monotonically rising
curve at least as far as $M_{v}=7$ ($V\simeq15.3$ in NGC
2516). There are relatively more bright stars in NGC 2516 than the
field because it is a very young cluster.
There is then, despite the correction for contamination, a 
significant minimum at $7.7<M_{v}<8.7$ ($16<V<17$), that coincides
with the ``Wielen dip'' \citep[see][]{wielen74, upgren81, bahcall86} -- 
that is also seen (at the level of 30-40 percent) in
the LFs of the field, the Pleiades and the young $\alpha$ Persei
cluster at $M_{v}\simeq7.5$
\citep{prosser92, meusinger96, belikov98, reidhawley00} 
and in the Hyades at $M_{v}=8.5$ \citep{reid92}. This dip is
defined with reasonable clarity in NGC 2516, due to the large numbers
of cluster members.  There is some evidence that the overall rise in
the LF towards fainter magnitudes then levels off for $M_{v}>9$ and
certainly there is no sign of the very steep rise in the LF of a factor
2-3 between $9<M_{v}<10$, that is seen in the Pleiades and nearby field
stars. In this respect, NGC 2516 is much more similar to the Hyades.

We \emph{do not} believe that incompleteness can be responsible for a
lack of faint stars in NGC 2516.  We have calculated that the LF is
complete to at least $M_{v}=11.7$, in the sense that stars are not
missed because they were not detected. Perhaps then the LF of NGC 2516
is different to that of the Pleiades and the field, but there are other
possibilities: First, it may be that that the contamination fraction is
increasingly overestimated for $V>17$ or underestimated for
$14<V<17$. We regard this as unlikely and in fact for the last column
in Table~\ref{contam}, we believe the contamination fraction is
probably \emph{underestimated} (see Sect.~\ref{contaminate}). Second,
we may have set our membership criteria too tightly and missed a
significant fraction of fainter cluster members that lie either above
or below our membership bounds in the $VI$ CMD. We already regard our
membership criteria as quite generous and thus the factor of
approximately two increase in the LF that would be required between
$10<M_{v}<12$, seems unlikely to be explained in this way.  Third, it
might be that mass segregation has been successful in removing lower
mass stars from the central cluster regions and hence our survey. This
has been put forward as an explanation of the differences between the
Hyades and field LFs \citep{reidhawley00} and although NGC 2516 is
much younger than the Hyades, we have only surveyed the central
regions. Mass segregation is discussed in more detail in
Sect.~\ref{masseg}. Lastly, it is possible that the {\em mass} function
in NGC 2516 is similar to that in the Pleiades, even though the LF is
not. This might be the case if the mass-luminosity relationship were
markedly different at low masses because NGC 2516 has a low metallicity
compared with the Pleiades. This possibility is discussed in Sect.~\ref{massfun}.

\subsection{Mass Function}

\label{massfun}

\begin{figure}
\vspace*{21.7cm}
\includegraphics{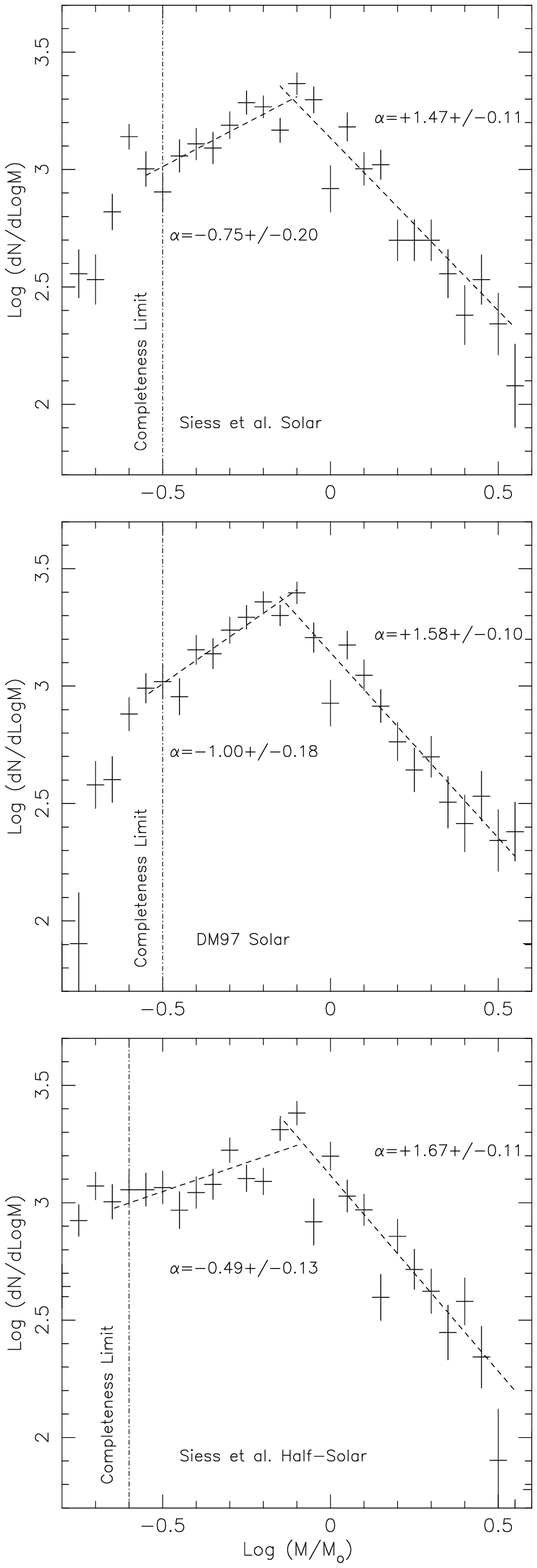}
\caption{
The corrected mass functions for NGC 2516 (see text), derived from the
solar metallicity model of \citet{siess00} (top), the solar metallicity
model of \citet{dantona97} (middle) and the half-solar metallicity
model of \citet{siess00} (bottom).
The dashed lines indicate the power law fits discussed in
Sect.~\ref{massfun} and the dash-dot lines indicate the approximate
mass at which our survey becomes incomplete.
}
\label{mfplot}
\end{figure}

The cluster MF can be explored in a preliminary way by calculating the
mass of each candidate member using a mass-colour relationship derived
from the isochrones we used to fit the cluster $VI$ CMD. Unresolved
binarity is of course a problem here.  The net result of assuming
that unresolved binaries are single stars, will be that the true MF
will be slightly steeper than that derived (in the sense that an
assumed single star of a given mass is actually two stars, one of which
could have a much lower mass -- see below). 
In principle, the numbers of low mass stars could be doubled
by hiding them in binary systems with more luminous companions.

The binarity in NGC 2516 and its effect on the derived MFs is discussed 
further in Sect.~\ref{binary}, but we note that
in Sect.~\ref{membership} we identified candidate members which are
approximately equal mass binary systems based on their position above
the cluster isochrone. Using the $BV$ CMD we can also identify
candidate binaries from among the bright stars added to the CTIO sample
from \citet{jeffriesn251697}. 
Each identified binary system is counted as if it consisted of two
identical components with the same colour. This is an approximation,
because for objects with mass ratios less than unity, one component
will be cooler. However, in the absence of better binarity information,
a more complex approach is not warranted.  Corrections for
contamination by non-members are applied by using the fractional
membership probabilities in Table~\ref{contam}. Each candidate will
contribute a fraction to the MF bin corresponding to its mass,
according to the value of $V-I_\mathrm{c}$ (or $B-V$
if $V-I_\mathrm{c}$ is unavailable) and whether it is a
candidate binary. 

The mass-colour relationships 
are derived from the same evolutionary models and empirical
isochrones used to select the cluster candidates in the first place.
The uncertain metallicity of NGC 2516 causes some difficulties
here. Stars of a given colour have smaller masses at lower
metallicities.  We perform our calculations using the solar 
and half-solar metallicity models of \citealp{siess00} and the solar
metallicity model of \citet{dantona97}, which were
discussed in Sect.~\ref{isochrone}.  Note that we
choose to use $V-I_\mathrm{c}$, rather than $B-V$, as the primary
indicator of $T_\mathrm{eff}$ and hence mass, because it is relatively
insensitive to metallicity \citep{leggett96} and covers a wider range
of masses.  Only the hottest stars have their masses calculated from
$B-V$, where $B-V$ is an almost metallicity-independent
$T_\mathrm{eff}$ proxy \citep{castelli99}, so the fact that we have
ignored the metallicity dependence of the $B-V$-$T_\mathrm{eff}$
relationship in deriving the low-metallicity isochrone should be
unimportant (see Sect.~\ref{isochrone}). 

The corrected MFs are shown in Fig.~\ref{mfplot}, expressed as
the number of stars per unit logarithmic mass interval, as a function
of logarithmic mass (in solar units). The approximate lower mass limit
at which our survey is complete (corresponding to $V=20$) is about
$M=0.3M_{\sun}$ for the solar metallicity models and $M=0.25M_{\sun}$
for the half-solar metallicity model.
In these diagrams, the canonical stellar initial MF of \citet{salpeter55}
would be a straight line of
the form $dN/d\log M \propto M^{-\alpha}$ with $\alpha=+1.35$. 

Our MF for NGC 2516 is well defined between 3$M_{\sun}$ (where any
uncertainty in the age of NGC 2516 would start to be a factor and also where small number
statistics become important) and 0.25--0.3$M_{\sun}$ (where
incompleteness sets in). Note that below the completeness limit we
cannot simply say that the points are lower limits because there is
also the uncertain level of contamination by cluster non-members to consider.
For each of the two metallicities, there is a clear Salpeter-like
rise in the MF as the mass decreases, followed by a peak at $M\simeq0.7M_{\sun}$ and 
a turnover towards lower masses. We have checked that the 
corrections for contamination by
non-members and for binaries have very little effect on this overall shape.

The exact form of the MF might of course be age, model and
metallicity-dependent. \citet{barrado01} investigated the
MF of M35, a rich northern hemisphere cluster of similar age to the
Pleiades and NGC 2516, using a variety of ages and (solar metallicity) evolutionary
models. Their results show that above $0.2M_{\sun}$, derived MFs are
virtually identical in all cases. 

We have parameterised our derived MFs in terms of two power law fits in
the range 0.3--0.7$M_{\sun}$ (or 0.25--0.7$M_{\sun}$ for the half-solar
metallicity model) and 0.7--3.0$M_{\sun}$.  For the higher mass range
we find $\alpha=+1.47\pm0.11$ and $\alpha=+1.67\pm0.11$ for the solar
and half-solar metallicity models of \citet{siess00}, and
$\alpha=+1.58\pm0.10$ for the solar metallicity model of
\citet{dantona97}. For the lower mass
range we find $\alpha=-0.75\pm0.20$ and $\alpha=-0.49\pm0.13$ for the
solar and half-solar metallicity models of Siess et al., and
$\alpha=-1.00\pm0.18$ for the solar metallicity model of D'Antona \&
Mazzitelli.  The
systematic difference between the models is caused by the change in
the colour-mass relationship. At lower metallicities, stars of the same
colour have lower masses, increasing the value of $\alpha$, but there
is also some modest model-dependence.
For simplicity of discussion in what follows we shall only use the two Siess et al.
models.  We {\em have} also done
all our calculations for the solar metallicity \citet{dantona97} model
and the results it yields in later sections are in reasonable agreement
with the Siess et al. solar metallicity model.

The power law fits to the higher mass range are in good agreement with
the Salpeter value, agree well with the average value of
$\alpha=+1.40\pm0.13$ found for intermediate mass stars in many open
clusters by \citet{phelps93} and are close to the ``universal'' field
initial MF of $\alpha=+1.3\pm0.3$ for $M>0.5M_{\sun}$ found by
\citet{kroupa01}. We also note that these quoted results neglect the
effects of binarity and used a single relationship between $V$ and mass
to calculate the MF from the LF.  This has the effect of making the MF
slightly less steep at high masses because binary systems are then
treated as one star with a slightly higher mass.  We have
investigated what difference this makes by simply treating our data in
the same way. If we were to adopt a single relationship between $V$ and
mass, irrespective of binary status, then the slopes of our derived
MFs would be smaller by about 0.2 and hence in even better agreement
with the previously published values.

Determinations of the MF in the Pleiades \citep{meusinger96} and M35
\citep{barrado01} clusters find MF slopes of $\alpha=+1.4$ and
$\alpha=+1.59\pm0.04$ for stars with $M>0.8M_{\sun}$. Meusinger et
al. then find that $\alpha\simeq 0$ in the Pleiades for
0.3--1.0$M_{\sun}$, whilst \citet{hambly99} estimate $\alpha=-0.3$ for
$M<0.5M_{\sun}$ in the Pleiades and Barrado y Navascu\'{e}s et
al. finds $\alpha=-0.2$ for M35 between 0.8 and 0.2$M_{\sun}$. The
latter two slopes take no account of binarity and should probably be
increased by \mbox{$\sim0.2$} (see above) 
before comparison with our results.  At very low
masses it is likely that the MF falls again.  \citet{bouvier98} and
\citet*{moraux01} find $\alpha=-0.4$ and $-0.5$ across the brown dwarf
boundary at $\sim0.05$--0.2$M_{\sun}$ in the Pleiades and Barrado y
Navascu\'{e}s et al. estimate a more extreme slope of $-1.8$ below
0.2$M_{\sun}$ in M35. This behaviour is mirrored in the field MF, where
a Salpeter-like slope is found above $0.6M_{\sun}$, a relatively flat MF 
with an $\alpha$ of $-0.1$ to $+0.3$ down to $\sim0.1M_{\sun}$, and then
a decline into the brown dwarf regime with $\alpha$ between $-0.5$
and $-1.0$ (\citealp*{gould97}; \citealp{reid99}; \citealp{chabrier00};
\citealp{kroupa01}).

If NGC 2516 has a solar metallicity, then its MF drops much
more sharply towards lower masses than occurs in either the Pleiades or
field for $0.3<M<0.7M_{\sun}$. This steep slope is
of course directly related to the deficit of low luminosity stars in
the NGC 2516 LF with respect to the Pleiades and the field which we
remarked upon in Sect.~\ref{lumfun}.
Any of the explanations offered there (such as mass-segregation) might make
the drop in the MF below 0.7$M_{\sun}$ less extreme. A Pleiades-like
MF could still be recovered for the solar metallicity scenario 
if the number of NGC 2516 members between 0.3 and 0.5$M_{\sun}$
were roughly doubled with respect to the more massive stars.
If NGC 2516 has a half-solar metallicity then the discrepancy with
the Pleiades and field MFs is reduced but still present.

\subsection{Binarity}

\label{binary}

\begin{figure}
\vspace*{6.3cm}
\includegraphics{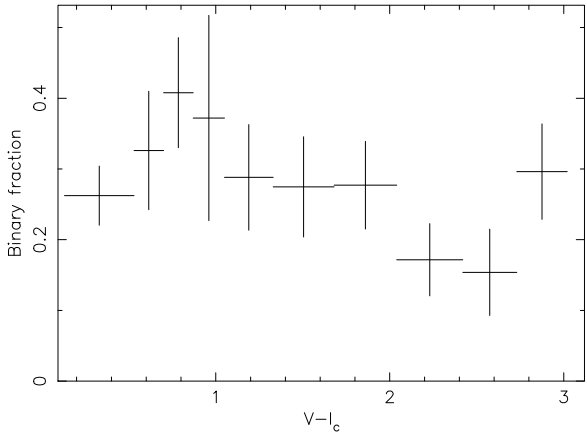}
\caption{Binary fraction in NGC 2516 for systems with approximately
$0.6<q<1$ (see Sect.~\ref{binary}) as a function of $V-I_\mathrm{c}$.
}
\label{binaryfrac}
\end{figure}

The frequency of binary systems and the distribution of their mass
ratios are important constraints on star formation theories.
So far in this work we have made a partial correction for binarity in
determining the cluster MF, by identifying cluster candidates with near
equal mass ratios ($q=M_{2}/M_{1}\simeq 1$). Unresolved binarity will also influence the
deduced total mass of the cluster (see Sect.~\ref{clustermass}). We have also not
considered the possibility of triple systems in this paper, which could
place an object up to 1.2 mag. above the cluster isochrone (for equal
mass components). However, the numbers of missed triple systems should be
quite rare -- less than 2 percent of systems in the Pleiades
\citep{mermilliod92}.

Figure~\ref{binaryfrac} shows the fraction of photometrically
identified binaries as a function of $V-I_\mathrm{c}$ colour in NGC
2516, where membership was 
determined from the solar metallicity isochrone of \citet{siess00}. 
The binary fraction is defined as the number of probable binary
systems divided by the total number of systems. The fractions are
calculated using our membership catalogue, the binarity flags and the
membership probabilities in Table~\ref{contam}. The weighted mean binary fraction
is $26\pm2$ percent with marginal evidence (at the 90 percent
confidence level) for a decrease in the binary fraction towards lower
masses. This may be more significant, as we expect the lowest mass bin
in Fig.~\ref{binaryfrac} to have an overestimated binary fraction,
because incompleteness sets in and will bias against single star
detection.  It is also the case that low mass M-dwarfs are more prone
to flaring, which might result in some fraction of the lowest mass
single stars being misclassified as binaries \citep[see][]{stauffer84}.  The
confidence level for a significant downward trend increases to 98
percent if this last point is excluded. On the other hand, the error in the binary
fraction should be increased to take into account the somewhat
arbitrary nature of the isochrones we have used to define the cluster
single star sequence. If we allow these isochrones to move up or down
by $\pm0.1$ magnitude (see Sect.~\ref{isochrone}), the mean binary
fraction changes by $\pm5$ percent. As we discussed in
Sect.~\ref{isochrone}, the shape of the half-solar metallicity
isochrone is slightly different for $V-I_\mathrm{c}>2.0$, which would
result in a differential increase in the binary fraction of cool stars
with respect to the hot stars of about 5 percent. This would be enough
to remove the possible trend just discussed, so it is premature to
claim to have seen any dependence of binary fraction on mass.

The binary fraction estimated in this way is a lower limit.
The mass ratios to which our photometric identification technique is
sensitive, can be estimated using the results and equations found in
\citet{kahler99}. We note that the $q$ sensitivity limit
\emph{is not} quite independent of colour for stars 0.3 mag above the single
star locus, because the gradient of the $V$ vs
$V-I_\mathrm{c}$ cluster isochrone is not constant. Using equation~5 in
\citet{kahler99} and the gradients determined from the cluster
isochrone we find that we are sensitive to $0.59<q<1.0$ for $V-I_\mathrm{
c}<0.6$, $0.55<q<1.0$ for $0.5<V-I_\mathrm{c}<1.1$, $0.62<q<1.0$ for
$1.1<V-I_\mathrm{c}<2.5$ and $0.60<q<1.0$ for $V-I_\mathrm{c}>2.5$. Given an
approximately flat distribution of $q$ or one that rises towards lower
$q$ values, any small variations in apparent binary fraction seen in
Fig.~\ref{binaryfrac} could also be partially explained by this varying
sensitivity.

In field stars, \citet{duquennoy91} found that the mass ratio
distribution increased towards lower mass ratios, perhaps peaking at
$q=0.3$, and that the binary fraction of solar-type stars was about 70
percent for all mass ratios, and 21 percent for $0.6<q<1.0$.  Work on
the Pleiades solar-type stars is in broad agreement with these results
(\citealp{mermilliod92}; \citealp*{bouvier97bin}). There is some
evidence that the binary fraction among lower mass field objects is
smaller (30-40 percent), but that they are more inclined to be found in
$q\simeq1$ systems \citep{fisher92, reidhawley00}.  Using
a similar photometric selection technique to the one used here,
\citet{stauffer84} found that 26 percent of Pleiades low mass stars
($0.7<V-I<2.1$) were more than 0.3 mag. above a single star isochrone
and that this was similar to the 22 percent photometrically determined
binary fraction in hotter stars found by \citet{bettis75}.

The main result of this subsection is that the binary fraction in NGC
2516 for stars with $q$ in the approximate range 0.6-1 is similar to
that of the Pleiades or the field, when the binaries are identified in
the same way.  If the distribution of $q$ were that proposed
for the field by \citet{duquennoy91}, down to $q=0$, then the
total binary fraction in NGC 2516 would be about 85 percent. If the
distribution of $q$ were flat, then the total binary fraction would be
65 percent.

A subsidiary consideration is estimating what effect unresolved
binarity might have on the derived mass functions.  Our identification
of $q>0.6$ binaries, where the companion star can significantly change
the colours and magnitudes of the system as a whole, results in MFs
which have slopes about 0.2 larger than those where this is ignored
(see Sect.~\ref{massfun}).  These MFs have the merit of beoing obtained from
observables which are in principle comparable with other estimates
derived from photometry data in other clusters, and do not rely on
knowledge of the mass ratio distribution. 
However, even for those systems with $q<0.6$ there can be a
significant number of ``hidden'' lower mass stars, resulting in a MF
{\it for all stars} which is even steeper still. \citet{sagar91} investigated
this effect in 2-14$M_{\sun}$ stars in LMC clusters. They found, for
{\it true} MFs with $\alpha=+1.5$, a binary fraction between 0.5 and 1,
and binary companions drawn randomly from the same MF, that the
observed MF (estimated from a mass-magnitude relation) had a slope that
was 0.3-0.4 smaller. The effect was even larger for smaller values of
$\alpha$. A true $\alpha$ of $+0.5$ might appear as $\alpha=-0.4$ for a
binary fraction of 1.0. Similar simulations by \citet{kroupa01} yield
differences between the system and single star MFs largely in agreement
with these results.

We point out to the reader that by using a colour-mass relation
and identifying and dealing with $q>0.6$ binaries, we have partially
alleviated this problem. To try an gauge by how much the true
stellar MF slopes might be further increased over the quoted MF
slopes in this paper, we randomly added binary companions to a fraction
of the ``single'' stars. Using a total binary fraction of 65 percent
and a flat $q$ distribution we find that the true single-star MF
$\alpha$ is increased by a further 0.05 for $0.7<M<3.0M_{\sun}$ and
0.3 for $0.3<M<0.7M_{\sun}$ (using the \citealp{siess00} solar
metallicity models). Using the $q$ distribution
proposed for field binaries by \citet{duquennoy91} and assuming a
total binary fraction of 85 percent, results in $\alpha$ increasing by
0.1 and 0.4 in these two mass ranges.

\subsection{The cluster mass}

\label{clustermass}

We can integrate our corrected MF (from the solar metallicity
isochrone) to yield an estimate of the total
mass of the cluster (down to about 0.3$M_{\sun}$) of 1105$M_{\sun}$. 
More than half of this mass is contained in stars with 0.6-2$M_{\sun}$
with decreasing contributions at lower and higher masses. 
The equivalent calculation for the MF derived from the half-solar metallicity
isochrone is 945$M_{\sun}$ (complete to 0.25$M_{\sun}$).

Unresolved binarity will increase the derived cluster mass.  We have
made a partial correction for this in our work so far, by identifying
near equal mass-ratio binary systems. If we had not done so, our
deduced cluster mass would have been only 880$M_{\sun}$ (solar
metallicity).  To estimate the maximum likely contribution that could
arise from binaries with $q<0.6$, we can use the $q$ distribution
proposed for field binaries by \citet{duquennoy91} and assume that the
total binary fraction is 85 percent and independent of primary
mass. Integrating this distribution, we find that the cluster would be
35 percent more massive than if all the stars were single, and thus the
total cluster mass would be 1190$M_{\sun}$ (solar metallicity), for primaries with
$M>0.3M_{\sun}$.  A correction could also be applied for stars less
massive than this. Integrating an extrapolated MF derived from the
\citet{siess00} solar metallicity isochrone (see Sect.~\ref{massfun})
from say 0.0-0.3$M_{\sun}$, yields only another 73$M_{\sun}$.  The
corresponding additional mass for the half-solar metallicity model is
$72M_{\sun}$ (0.0--0.25$M_{\sun}$).  Thus irrespective of the
cluster metallicity, the contribution of low-mass stars and brown
dwarfs to the total cluster mass \emph{inside our surveyed area} is
likely to be less than 10 percent unless there were a sharp upturn in
the MF below 0.3$M_{\sun}$.

The cluster mass is similar, but a little higher, than that found for
the Pleiades in the same mass range. \citet{meusinger96} quote a
figure of 800$M_{\sun}$ for $M>0.3M_{\sun}$ and \citet{pinfield98} derive
735$M_{\sun}$ for all masses down to the substellar limit. However, we
have to be careful to compare like with like. The quoted Pleiades
results are extrapolations for all stars out to the cluster tidal radius.

\citet{pinfield98} give an approximate expression for the tidal radius of
a cluster in a circular Galactic orbit, close to the Sun as
\begin{equation}
r_{t} = \left(\frac{GM_{c}}{2(A-B)^{2}}\right)^{1/3}\, ,
\label{tidal}
\end{equation}
where $M_{c}$ is the cluster mass and $A$ and $B$ are the Oort
constants in the solar neighbourhood. Using the value of $A-B$ from
\citet{kerr86}, this reduces to $r_{t}=1.46M_{c}^{1/3}$,
with $M_{c}$ in solar masses and $r_{t}$ in parsecs. 
If we take 1000$M_{\sun}$ as the minimum mass for NGC 2516, $r_{t}\geq 14.6$\,pc.

Using a distance modulus of 7.9, our survey of the cluster covers a
square area, 6.2\,pc on a side. Thus we expect significant numbers of
cluster members outside the area covered by our
survey. \citet{pinfield98} 
find that the Pleiades has $r_{t}=13.1$\,pc and that
although the high mass stars are highly concentrated within a few pc of
the cluster centre, there are significant numbers of low mass stars at
much greater distances, such that only half the cluster mass was
contained within 3.66\,pc.  If NGC 2516 is analogous to the Pleiades in
terms of the spatial distribution of its members and the amount of mass
segregation present (see Sect.~\ref{masseg}), then the total cluster mass
might be significantly greater than just what we have observed in our limited survey.

\subsection{Mass segregation}

\label{masseg}

\begin{figure}
\vspace*{18.5cm}
\includegraphics{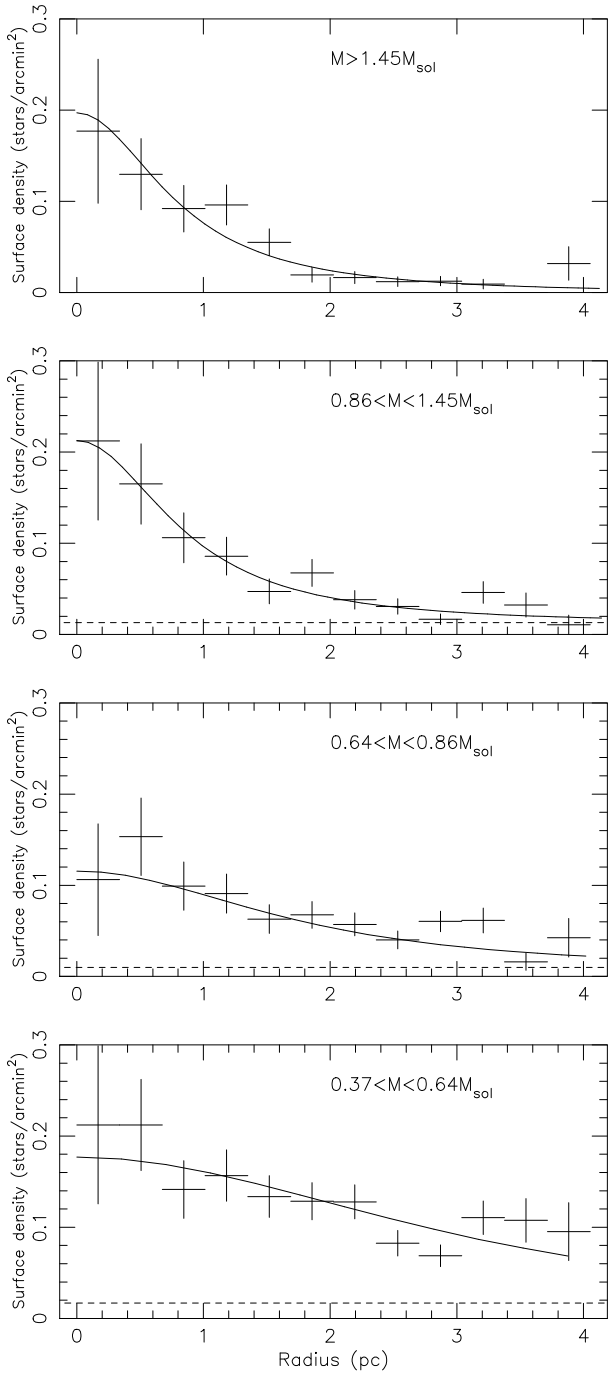}
\caption{The radial distribution of stellar surface density in NGC
2516, split into 4 approximate mass bins (see text). The x-axis values
assumes a cluster distance modulus of 7.9. The horizontal dashed lines
indicate the level of (assumed uniform) contamination expected in each
of these samples. The solid lines are King profile
fits for assumed tidal radii of 14.6\,pc (see Table~\ref{kingfits}).
}
\label{kingplot}
\end{figure}

Dynamical evolution and mass segregation can have a significant effect
on the shape of the present day LF and MF of NGC 2516. Equipartition
leads to a growing core radius with decreasing mass and the
preferential evaporation of low-mass cluster members (see for example
\citealp{delafuente95}; \citealp{kroupa95}; \citealp{delafuente00}; 
\citealp*{kroupa01b}). The evidence for
dynamical effects on the MFs of open clusters older than a few hundred Myr
is strong (e.g. \citealp{reid92}; \citealp*{montgomery93}; \citealp{sarajedini99}).
The segregation
effects are less obvious, but still seen, in younger clusters like the
Pleiades and M35 \citep{pinfield98, barrado01},
with a hint that equipartition has not been achieved in the lowest mass
stars, perhaps because these clusters have ages similar to their
dynamical relaxation timescales.

As a step towards understanding how much mass segregation might have
taken place in NGC 2516, we examine the radial surface density profiles
of candidate cluster members 
split up into bins according to their $V-I_\mathrm{c}$ colours
and hence masses. We exclude binary candidates at this stage, because
their colours are not reliable mass indicators -- they could be up to
twice as massive.  We have to acknowledge
though that we must have included some binary stars with $q<0.6$, which
will have the effect of blurring any distinction between different mass
bins, perhaps lessening the effects of mass segregation.  In
order to have sufficient stars to draw statistically significant
conclusions we divide the stars into just 4 mass bins (see Table~\ref{kingfits}). 
The division is made preserving the colour
boundaries in Table~\ref{contam}, so that estimates of field star contamination
can still be made. We do not include the lowest mass interval from
Table~\ref{contam} because of possible uncertainties in the exact level of
contamination (see Sect.~\ref{contaminate}). Note that the values for
the boundaries of the mass bins are derived from the solar metallicity
\citet{siess00} isochrone. The corresponding mass boundaries in
Table~\ref{kingfits} for the half-solar metallicity would be:
1.38$M_{\sun}$, 0.79$M_{\sun}$, 0.58$M_{\sun}$ and 0.27$M_{\sun}$.

The centre of mass for each bin is determined by minimising the projected
moment of the stars about a point. The centres drift south-west by
about 2.5 arcminutes between the highest and lowest mass bins. This small
drift has no effect on our results, which would be almost identical if
we had fixed the cluster centre at a single position. The surface
density of stars as a function of radius (assuming a cluster distance
modulus of 7.9) is then calculated for each mass bin, taking account of the
geometry of the surveyed area.

The results are shown in Fig.~\ref{kingplot}. The horizontal dashed
line in each plot indicates the (assumed uniform) density of
contaminating objects.  It now becomes clear that the probability that
a candidate member is actually a contaminating object shrinks
drastically if we only consider objects within $\sim15$ arcminutes
($\sim1.7$\,pc) of the cluster centre. Conversely, if we consider
objects in the outer part of our survey, the probability of membership
is lower than the average values in Table~\ref{contam}.

\begin{table*}
\caption{King profile fitting results (assuming a cluster distance
modulus of 7.9). Columns 1 and 2 show the mass subset considered (for the
solar metallicity isochrone -- see text), col. 3 is
the assumed tidal radius, col. 4 the fitted core radius, col. 5 the
fitted normalisation constant, col. 6 the expected surface density of
contaminating field stars and col. 7 the $\chi^{2}$ of the fit (with 10
degrees of freedom). For each mass subset, cols. 8 and 9 list the 
total numbers of cluster members in our survey and the total number of
cluster members out to the tidal radius predicted by the fitted model.}
\begin{tabular}{cccccccccc}
$V-I_\mathrm{c}$ & Mass  & Centre & $r_{t}$ & $r_{c}$ & $\rho_{0}$ & $\rho_\mathrm{
cont}$ & $\chi^{2}$ & $N_\mathrm{ survey}$ & $N_\mathrm{ tidal}$ \\
subset  & ($M_{\sun}$) & (J2000.0)& \multicolumn{2}{c}{(pc)} & 
\multicolumn{2}{c}{(stars/arcmin$^{2}$)} & & & \\  
&&&&&&&&&\\
$<0.53$ & $>1.48$ & RA = 7 58 03.2 & 14.6 & $0.84^{+0.17}_{-0.14}$ & 0.223 & 0.0 & 7.7 &
92 &  113 \\
 & & Dec = -60 45 29 & 18.4 & $0.81^{+0.16}_{-0.14}$ & 0.218 & 0.0 & 7.8 &
92 &  121  \\
$0.53-1.05$ & $0.87-1.48$ & RA = 7 58 05.6 &14.6 & $0.90^{+0.23}_{-0.17}$ & 0.228 &
0.013 & 10.9 & 103 & 126 \\
 & & Dec = -60 46 48 &18.4 & $0.86^{+0.21}_{-0.16}$ & 0.225 &
0.013 & 11.0 & 103 & 135 \\
$1.05-1.68$ & $0.65-0.87$ & RA = 7 57 56.0 &14.6 & $1.91^{+0.56}_{-0.40}$ & 0.141 &
0.010 & 11.9 &  150 & 196 \\
 & & Dec = -60 47 00 &18.4 & $1.80^{+0.53}_{-0.37}$ & 0.132 &
0.010 & 11.9 &  150 & 216 \\
$1.68-2.73$ & $0.35-0.65$ & RA = 7 57 48.0 &14.6 & $3.52^{+0.99}_{-0.64}$ & 0.279 & 0.017 &
13.4  & 280 & 637 \\
 & & Dec = -60 47 00 & 18.4 & $3.26^{+0.87}_{-0.56}$ & 0.238 &
0.017 & 13.1 & 280 & 725 \\
\end{tabular}
\label{kingfits}
\end{table*}

To get some parameterisation of the mass segregation we have 
fitted empirical profiles of the form \citep[see][]{king62}
\begin{equation}
\rho = \rho_{0}\, \left( \frac{1}{\sqrt{1 + (r/r_{c})^{2}}} - 
\frac{1}{\sqrt{1 + (r_{t}/r_{c})^{2}}} \right)^{2}\, ,
\label{profile}
\end{equation}
where $\rho_{0}$ is a normalisation constant, $r_{c}$ is the ``core
radius'', where the surface density falls to half its central value,
and $r_{t}$ is the tidal radius where the cluster is truncated. 
The King
model assumes that there is no contamination in the sample, so we
subtract the levels indicated by the dashed lines in Fig.~\ref{kingplot}
before fitting. Since our data are confined to regions well within
$r_{t}$, the fits are relatively insensitive to this parameter. To test 
this we performed two sets of fits, fixing $r_{t}$
at either 14.6\,pc (corresponding to a minimum cluster mass of 1000$M_{\sun}$
-- see equation~\ref{tidal}) or 18.4\,pc for a cluster mass of
2000$M_{\sun}$, which acknowledges that a significant fraction of the
cluster mass may lie outside the CTIO survey.

The fits are all reasonable, with $\chi^{2}$ values of between 7.7 and
13.4, with 10 degrees of freedom in each case. The fit results,
including 68 percent confidence intervals in $r_{c}$, and the assumed
surface density of contaminants, are reported in Table~\ref{kingfits}.

The King profile modelling suggests that the core radius undergoes a
dramatic enlargement as the mass decreases from $\sim1.0M_{\sun}$ to
$\sim0.5M_{\sun}$ and that this is independent of the assumed value of
$r_{t}$. It could be argued that this might be explained in terms of an
underestimation of the contamination level in the two lower mass bins. Such
an underestimation would lead to a ``flatter'' distribution of surface
density that would then be fitted with a large value of $r_{c}$. To
test this, we re-fitted the two lowest mass bins with a King profile
plus a \emph{variable} uniform surface density, which simulates an extra
unaccounted for level of contamination.  The core radius was fixed at
0.9\,pc, corresponding to the value of $r_{c}$ for stars with mass
$0.86<M<1.48M_{\sun}$. We found that the $\chi^{2}$ values were
similar with this model ($\chi^{2}$ values of 13.6 and 11.5
respectively for the same numbers of degrees of freedom), but the
levels of contamination implied by the fits (i.e. the extra constant
surface density terms) were factors of 3--5 higher than we estimated in
Sect.~\ref{contaminate}. This would mean that our estimates of the
contamination levels were wildly in error and that only about 35
percent of the cluster candidates with $1.05<V-I_\mathrm{c}<2.73$ were
members in contrast to our present estimate of 84 percent. We do
not believe that errors of this extent are possible.

We conclude that the evidence for mass segregation is strong when
comparing stars above and below about 0.8$M_{\sun}$.  This coincides
with the break in slope of the MF and lead us to suspect that
the downturns in the LF and MF of NGC 2516 at low masses may be
explained by mass segregation (see Sects.~\ref{lumfun} and
\ref{massfun}). We can test this by
extrapolating and integrating the surface density profiles beyond the
extent of the current survey. \citet{king62} gives an expression for the
integral of equation~\ref{profile} out to the tidal radius, which we use to
calculate the total number of cluster stars in each mass bin, $N_\mathrm{
tidal}$, in Table~\ref{kingfits}. This number can be compared with the actual number
of cluster members (corrected for contamination), $N_\mathrm{ survey}$, to estimate what
fraction of cluster stars lie outside our survey. Naturally, such an
extrapolation is reliant on a knowledge of the surface density
profile \emph{outside} the area we have surveyed!

The fraction of cluster stars \emph{inside} our survey decreases from 
81 percent for $M>1.48M_{\sun}$ to 44 percent for
$0.37<M<0.64M_{\sun}$, if $r_{t}=14.8$\,pc. 
These fractions decrease to 76 percent and 39 percent for the larger
value of the cluster mass and $r_{t}$. 

There are two major implications of this result. (1) The total cluster
mass is probably greater than we first estimated.  Most of the cluster
mass is concentrated in stars between 0.6--2$M_{\sun}$ so it is likely
that the cluster is more massive by about a factor of 1.3.  (2) The
decrease in the MF towards lower masses will be flattened
off. Earlier we stated that an increase of roughly a factor of two in
the numbers of stars below 0.5$M_{\sun}$ (with respect to the numbers
of more massive stars) could bring the LF of NGC 2516 roughly
into agreement with that of the Pleiades. The results in Table~\ref{kingfits} show
that within our survey area the ratio of stars with
$0.37<M<0.64M_{\sun}$ to those with $M>1.48M_{\sun}$ is 3.0, but that
this ratio for \emph{all} cluster stars is probably as large as 5.6-6.0;
approximately the increase required. Taking the results from the lowest
two mass bins, and assuming that the correction to the MF will vary
linearly (with log mass) from about 1.3 at 0.7$M_{\sun}$ to 2.3 at 0.3$M_{\sun}$,
the fitted values of $\alpha$ (see Sect.~\ref{massfun}) 
would increase from $-0.75\pm0.20$ to
$-0.08\pm0.19$ and from $-0.49\pm0.13$
to $+0.13\pm0.13$ for the solar and half-solar metallicity models of
\citet{siess00} respectively.

It is difficult to provide precise corrections to the MF and total
cluster mass because our analysis is dependent to a great extent on the
assumed form of the surface density distribution {\em outside} of the region
surveyed and to a lesser extent on the assumed non-member contamination fraction in each
bin. For example, if the level of contamination for low-mass stars is
higher than we have assumed then: the numbers of low mass stars inside
our survey will be smaller; the fitted core radius will decrease and
the fraction of cluster members inside the survey will increase; the
cluster mass inside the survey will be smaller; the MF will drop more
steeply and the total cluster mass will be smaller. To give an idea of
the possible size of these effects we ran through our analysis again,
but systematically increased the number of contaminants predicted in
rows 3 and 4 of Table~\ref{contam}, for the columns with $V-I_\mathrm{c}>1.68$
($M<0.64M_{\sun}$), by \emph{twice} their Poisson errors. The results
from this were: that the MF slope below 0.6$M_{\sun}$ changed from
$\alpha=-0.75$ to $\alpha=-0.92$; the cluster mass inside the survey
decreased slightly from $1105M_{\sun}$ to 1050$M_{\sun}$; the core
radius in the lowest mass bin decreased from 3.5\,pc to 2.8\,pc; and
the fraction of stars with $0.37<M<0.64M_{\sun}$ inside our survey
increased from 44 to 52 percent. Changes of this order do not affect
any of our major conclusions.

\section{Discussion}

\label{discussion}

NGC 2516 has been referred to as the ``southern Pleiades'' on the basis
of their similar ages and apparent richness of high mass stars \citep{eggen83a}.
The central questions which we can attempt to answer on the
basis of the results presented here are: (1) how similar are NGC 2516
and the Pleiades, in terms of their LFs, MFs, total mass, mass
segregation and binarity, especially among low-mass stars? (2) What
are the prospects for finding even lower mass objects and brown dwarfs
in NGC 2516?

\subsection{NGC 2516 and the Pleiades}

Of key importance in any comparison with the Pleiades is to make sure
we are comparing similar surveys, in terms of their area and
completeness. We have seen that our NGC 2516 survey is complete to $V=20$,
($M_{v}\simeq 11.7$) and covers the central
$6.2\times6.2$\,pc. The Pleiades is approximately 3 times closer
and thus surveys of the Pleiades must cover 9 times the area of our NGC
2516 survey to be equivalent.

\citet{dachs89} presented the LF for bright stars ($M_{v}<5$) in
NGC 2516, from a 1 degree diameter central region. They showed that this
LF was similar in shape to that of the Pleiades with approximately a
factor of two more early-type stars in NGC 2516. There was marginal
evidence for mass segregation in their sample, but few stars at this
luminosity were estimated to lie outside their survey. On this basis
they claimed that NGC 2516 was about twice as massive as the Pleiades,
implicitly assuming that the two clusters have similarly shaped MFs 
at all masses.

The Pleiades survey of \citet{meusinger96} covered the central 16.5 square degrees
(about $9.3\times9.3$\,pc) and was complete to $M_{v}=12$. The LF shows a
relatively sharp increase of a factor $\sim2$ at $M_{v}=10$ and the MF
is quite flat, with $\alpha\simeq0$ (see Sect.~\ref{massfun}),
between $0.3<M<1.0M_{\sun}$. The LF and MF were simply corrected by a
linear factor for stars of all masses and luminosities to account for
Pleiades stars which lay outside the survey area. Thus no account of
mass segregation was taken other than a crude correction to the cluster
mass, which resulted in $M_\mathrm{ Pleiades}=800M_{\sun}$ down to stars
of $0.3M_{\sun}$. \citet{hambly99} derived the MF from a wider 36
square degree survey ($13.7\times13.7$\,pc) and obtained $\alpha\simeq
-0.3$ below 0.5$M_{\sun}$.

Using survey material complete to radii of about 6\,pc, \citet{pinfield98}
show that core radii in the Pleiades increased from about
1\,pc for stars of 3-5$M_{\sun}$ solar masses to 3\,pc at
0.3$M_{\sun}$.  As a result we can say that Meusinger et al.'s work
must have been affected by mass segregation, with roughly one third of
the low-mass members lying outside their survey, but essentially all
the Pleiades stars with $M>1.2M_{\sun}$ included.  Hambly et al.'s
work fares better, with perhaps only $\sim20$ percent of stars
with $M<0.5M_{\sun}$ missing. In both cases then, the true value of $\alpha$
is likely to be a little higher than derived from these spatially
limited surveys, although the effect should not be nearly as large as
calculated for NGC 2516 in Sect.~\ref{masseg}.
Pinfield et al.  estimate that the total
mass of the Pleiades, after accounting for this segregation is
735$M_{\sun}$.

The core radius values in the Pleiades are comparable with the NGC 2516
values that we determined in Table~\ref{kingfits}, but it does seem that the growth
of $r_{c}$ with decreasing mass is more rapid in NGC 2516. A simple
power-law fit indicates that $r_{c}\propto M^{-\beta}$ with
$\beta\simeq 0.9\pm0.2$, rather than the 0.5 found for the Pleiades by
\citet{pinfield98}. In a virialised
system, we might expect $r_{c}\propto M^{-1/2}$ (see
\citealp{pinfield98} for an analytic argument and \citealp{spitzer75}
for numerical simulations). This is perhaps a hint
that at least some of the mass segregation in NGC 2516 is
primordial, as has been supposed for some younger clusters
\citep{sagar88,bonnell98}, with the high-mass stars being initially more centrally
concentrated than the low-mass stars. 
This would be a mildly surprising result, because
dynamical mass segregation should have removed the signature of initial
conditions on the cluster relaxation timescale -- which is about 100\,Myr
for clusters of the size of NGC 2516 and the Pleiades.

Both the total mass and the MF of NGC 2516 are dependent on what is
assumed for the metallicity of the cluster and also to a lesser extent
on which stellar evolution models are used. If the cluster has a solar
metallicity then the mass inside our survey area (for stars with
$M>0.3M_{\sun}$) is
1100--1200$M_{\sun}$, depending on how unresolved binarity is
treated. The equivalent figure for a half-solar metallicity model is
950--1050$M_{\sun}$ (for stars with $M>0.25M_{\sun}$). 
If the cluster does have a solar metallicity then
the MF drops sharply below 0.7$M_{\sun}$, with $\alpha\simeq -0.75$. A
half-solar metallicity yields a shallower slope of $\alpha\simeq -0.49$.
We have established that most high mass ($>1.5M_{\sun}$) stars are
included in our survey, but that more than half of the low mass
($<0.6M_{\sun}$) stars of the cluster may lie outside this region --
depending on the exact form of the density distribution beyond our
surveyed area. Correcting the total cluster mass for this segregation
leads to an estimate of about 1240--1560$M_{\sun}$ -- twice the mass of the
Pleiades, in agreement with the earlier prediction made by
\citet{dachs89}.  This implies that the shapes of the MFs of NGC 2516
must be reasonably similar (at least over the mass range which
contributes most to the total cluster mass). As we discussed in
Sects.~\ref{massfun} and \ref{masseg}, a factor of two increase in the
numbers of low mass stars relative to high mass stars could increase
$\alpha$ at low masses and bring the NGC 2516 MF into into
agreement with the flat ($\alpha\simeq 0$) Pleiades MF between
$0.3<M<1.0M_{\sun}$. Mass segregation appears to provide just
this correction, increasing $\alpha$ by about 0.7.

Recent simulations of an evolving star cluster
by \citet{kroupa01b} show both the total MF and the MF for stars within
2\,pc of the centre of a cluster similar in size to the Pleiades, and
at an age of 100\,Myr. Their Figures~14 and~15, demonstrate that mass
segregation can indeed produce a downturn in the MF of the central regions below
0.7$M_{\sun}$, even if the whole-cluster MF is flat.  A thorough survey
for low-mass stars outside the area discussed in this paper will be
vital to constrain the surface density of members, the whole-cluster
mass function and the total cluster mass.

On the basis of a comparison of high mass ratio ($q>0.6$), unresolved
binary systems, the Pleiades and NGC 2516 have a similar binary
fraction of $26\pm5$ percent. It remains to be seen whether this similarity
persists for lower mass ratio binary systems.

In summary, subject to a wider survey confirming the presence of
an extended population of low-mass objects in NGC 2516 (see the offset
field of Hawley et al. 1999 for some preliminary evidence), we believe
that NGC 2516 and the Pleiades are similar in most respects, but that
NGC 2516 has twice the mass and numbers of stars.

\subsection{Lower mass stars and brown dwarfs in NGC 2516?}
\label{bddiscussion}

Any \emph{apparent} difference between the Pleiades and NGC 2516 MFs at
low masses could be very significant for those searching for even lower
mass stars and BDs in NGC 2516.  If we take our observed NGC 2516 solar
metallicity MF below $0.7M_{\sun}$, with $\alpha=-0.75$ (see
Sect.~\ref{massfun}), and simply extrapolate to lower masses, we find
that there would only be about 100 brown dwarfs (with
$0.030<M<0.075M_{\sun}$) in the area we have surveyed.  The half-solar
metallicity MF, with $\alpha=-0.49$, predicts around 220.
The total number of brown dwarfs in the cluster could be much larger if
mass segregation continues to lower masses. At the very least these
numbers should be doubled to account for the segregation that is
already apparent in $\sim0.5M_{\sun}$ stars. 

If instead we assume a Pleiades-like MF that is flat between 0.7$M_{\sun}$
and 0.2$M_{\sun}$ and then decreases with $\alpha=-0.5$ below this, we
calculate that there would be 360--440 brown dwarfs in the solar and
half-solar metallicity cases respectively.  Even with this MF, the total mass in
the form of brown dwarfs would be less than 3 percent of the cluster
mass and the extra contribution in the form of stars with $M<0.3M_{\sun}$
would be about 15 percent.  A search for brown dwarfs and low mass
stars in the outer regions of NGC 2516 will tell us much about the
whole-cluster mass function and the cluster dynamics.

\section{Summary}

In this paper we have presented a large, accurate and uniform
$BVI_\mathrm{ c}$ survey of stars covering 0.86 square degrees ($\simeq
6.2\times6.2$\,pc) of NGC 2516, and which is almost complete to $V=20$,
$M\simeq0.3M_{\sun}$. Thanks to the relatively low contamination of the
cluster main sequence by foreground and background objects we have been
able to select a sample of candidate cluster members. NGC 2516 probably
has a lower metallicity than the similarly aged Pleiades and our list
of candidate members list will be an important source of
\emph{optically} selected low-mass targets for further investigations of
magnetic activity, elemental abundances, lithium depletion and rotation
rates in convective stars.

We have made a preliminary investigation of the luminosity function,
mass function, binarity, mass segregation and total mass of NGC 2516, and compared
our results with the Pleiades, other young clusters and the
field. Because the metallicity of NGC 2516 may be as low as half the
solar value, we have performed two sets of calculations, one for a mass
fraction of heavy elements, $z=0.02$ and another with $z=0.01$.
Some of our results are somewhat dependent on the assumed metallicity and others
depend on extrapolating to lower masses or larger spatial areas. 
We can summarize the results of our investigation as follows:
\begin{enumerate}

\item
The luminosity function of the central regions of NGC 2516
are consistent with the Pleiades and field populations for
$M_{v}<8$. For fainter stars, the luminosity function of NGC 2516
flattens and contains a factor of two fewer stars than are seen in the
Pleiades and field.

\item The derived mass function for NGC 2516 is in agreement with the
Pleiades and field for stars with $0.7<M<3.0M_{\sun}$, 
with some metallicity and evolutionary model dependence. We find a mass
function, $dN/d\log M \propto M^{-\alpha}$, 
with $\alpha=+1.47\pm0.11$ and $\alpha=+1.67\pm0.11$ for the
solar and half-solar metallicity models of \citet{siess00} and
$\alpha=+1.58\pm0.10$ for the solar metallicity models of
\citet{dantona97}. These values are close to the
canonical \citet{salpeter55} value of $\alpha=+1.35$.
At lower masses ($0.3<M<0.7M_{\sun}$) 
we find $\alpha=-0.75\pm0.20$ and $\alpha=-0.49\pm0.13$
for the solar and half-solar metallicity Siess et al models, and
$\alpha=-1.00\pm0.18$ for the solar metallicty D'Antona \& Mazzitelli
models. These are more steeply
declining functions of mass than seen in either the Pleiades or field.

\item
Mass segregation is clearly present in NGC 2516 when we compare the
radial distributions of stars above and below about $0.8M_{\sun}$.
From extrapolation of simple analytic models for the surface density
distribution, we infer that about half of the lower mass cluster members
lie outside our surveyed area, whilst the vast majority of higher mass
stars are included. If this is confirmed by wider surveys for cluster
members, then the whole-cluster mass and luminosity 
functions of NGC 2516 and the Pleiades could be
very similar at least down to 0.3$M_{\sun}$. The cluster core radius increases
towards lower masses at a faster rate than if the segregation were
simply due to equipartition. This may indicate some remnant primordial
mass segregation in NGC 2516, where the high-mass stars were initially more
centrally concentrated than the low-mass stars.

\item
The binary fraction of A to M-type stars, with mass ratios of 0.6--1,
is $26\pm5$ percent in NGC 2516, which is comparable with the same
statistic for the Pleiades and field stars. This is a lower limit to
the total binary fraction, which could be as high as 65 to 85 percent,
depending on what form the mass ratio distribution takes. Unresolved
binarity has only been partially taken onto account in our mass
function determinations, in the sense that we {\em have} identified
those systems composed of two stars of roughly equal mass and have not
treated them as a single star with slightly higher mass. Mass functions
determined in this way at least have the merit of comparability with estimates
derived from photometric data on other clusters in the literature, but
we recognize that the slopes of the true stellar mass functions may be
a little steeper ($\alpha$ increases by $\sim0.05-0.1$ for high mass stars
and $\sim0.3-0.4$ for lower mass stars), because low mass stars could still
be hidden in many systems.

\item
The total mass of the cluster for stars with $M>0.3M_{\sun}$ is at
least $950M_{\sun}$ and probably as high as $1200M_{\sun}$,
depending on the metallicity, binary fraction and mass-ratio
distribution. Correcting for the fraction of stars that lie outside our
survey, but within the likely cluster tidal radius, may increase this
estimate to around 1240--1560$M_{\sun}$. NGC 2516 is therefore about twice
as massive as the Pleiades.

\item
If the whole-cluster mass functions of NGC 2516 and the Pleiades are
similar, then we expect about 360--440 brown dwarfs in NGC
2516. Extrapolation of the mass functions derived from our data suggest
that 100--220 would be in the area surveyed in this paper.

\end{enumerate}

\section*{Acknowledgements}
We would like to thank the director and staff of the Cerro Tololo
Interamerican Observatory, operated
by the Association of Universities for Research in Astronomy, Inc.,
under contract to the US National Science Foundation.  RDJ and MRT acknowledge
the financial support of the UK Particle Physics and Astronomy Research
Council (PPARC) throughout a large proportion of this work.
Computational work was performed on the Keele, Birmingham and Edinburgh
nodes of the PPARC funded Starlink network.

\bibliographystyle{apj}
\bibliography{iau_journals,master}

\label{lastpage}
\end{document}